\documentclass[journal=jpca,manuscript=article]{achemso}
\SectionNumbersOn

\usepackage{graphicx}% Include figure files
\usepackage{dcolumn}% Align table columns on decimal point
\usepackage{bm}% bold math
\usepackage[utf8]{inputenc}
\usepackage[T1]{fontenc}

\usepackage{tabularx}
\usepackage{mathptmx}
\usepackage{graphicx}
\usepackage{dcolumn}
\usepackage{braket}
\usepackage{xfrac}
\usepackage{hyperref}
\usepackage{tikz-cd}
\usepackage{float}
\usepackage{amsmath}
\usepackage[nameinlink,capitalise]{cleveref}
\usepackage{mathtools}

\usepackage{amsfonts}
\usepackage{amssymb}
\usepackage{braket}

\usepackage{multirow}
\usepackage{epstopdf}
\usepackage[normalem]{ulem}

\usepackage{hyperref}
\hypersetup{
	colorlinks=true,
	linkcolor=blue,
	citecolor=blue,
	filecolor=green,
	urlcolor=blue,
}

\newcommand{\matr}[1]{\mathbf{#1}}

\makeatletter
\renewcommand*{\acs@tocentry@print@aux}{%
	\begingroup
	\let\@startsection\acs@startsection@orig
	\acs@section*{\tocentryname}%
	\tocsize
	\sffamily
	\singlespacing
	\begin{center}
		\begin{minipage}{\acs@tocentry@height}
			\vbox to \acs@tocentry@width{\acs@tocentry@text}%
		\end{minipage}%
	\end{center}%
	\endgroup
}
\makeatother

\title{\Large General classification of qubit encodings in ultracold diatomic molecules}

\author{Kasra Asnaashari$^1$, Roman V. Krems$^1$, and Timur V. Tscherbul$^{2}$}
\affiliation{$^{1}$Department of Chemistry, University of British Columbia, Vancouver, V6T 1Z1, Canada \\
	$^{2}$Department of Physics, University of Nevada, Reno, NV, 89557, USA}

\email{ttscherbul@unr.edu}

\begin{document}

	%\begin{tocentry}
	%\includegraphics[scale=1.5,  trim = 30 0 0 0]{Fig_TOC.pdf}.
	%\end{tocentry}
	
%	\date{\today}
	
	\begin{abstract}
		Owing to their rich internal structure and significant long-range interactions, ultracold molecules have been widely explored as carriers of quantum information. Several different schemes for encoding qubits into molecular states, both bare and field-dressed, have been proposed. At the same time, the rich internal structure of molecules leaves many unexplored possibilities for qubit encodings. We show that all molecular qubit encodings can be classified into four classes by the type of the effective interaction between the qubits. In the case of polar molecules, the four classes are determined by the relative magnitudes of matrix elements of the dipole moment operator in the single molecule basis.  	We exemplify our classification scheme by considering a new type of encoding of the effective spin-1/2 system into non-adjacent rotational states (e.g., $N=0$ and $N=2$) of polar {and non-polar} molecules with the same nuclear spin projection. Our classification scheme is designed to inform the optimal choice of molecular qubit encoding for quantum information storage and  processing applications, as well as for dynamical generation of many-body entangled states and for quantum annealing.
		%The proposed classification scheme may also prove useful in the search of new molecular qubits, and in evaluating the relative merits of different qubit encodings. 

	\end{abstract}
	\maketitle

	\newpage
	\section{Introduction}

	Recent experimental progress towards high-fidelity quantum control of  ultracold molecules trapped in optical lattices \cite{Bohn:17} and tweezers \cite{Kaufman:21}  has stimulated much interest in using ultracold molecular gases for quantum information science (QIS) applications. The key advantage offered by ultracold molecules lies in their numerous and diverse  degrees of freedom, which include not only electronic and hyperfine states (which are also present in atoms), but also vibrational and rotational modes, all of which could be used to encode a qubit.  Additionally, these degrees of freedom allow one to encode quantum information into higher-dimensional Hilbert spaces, which could be used either for high-dimensional quantum computing \cite{Sawant:20} or  quantum error correction \cite{Albert:20}. Another crucial advantage of ultracold polar molecules for QIS is afforded by their strong, anisotropic, and tunable electric dipolar (ED) interactions, which can be used to engineer quantum logic gates \cite{DeMille:02,Yelin:06,Ni:18} and to generate many-body entangled states \cite{Micheli:06,Bilitewski:21,Tscherbul:23}.

	Several types of molecular qubit encodings have been proposed, including into adjacent rotational states  ($N=0$ and 1) of the same nuclear/electron spin projection \cite{DeMille:02,Yelin:06}, nuclear/electron spin sublevels of a single ($N=0$) rotational state \cite{Yelin:06,Park:17,Gregory:21,Herrera:14,Karra:16}, nuclear/electron spin sublevels of  adjacent  rotational states ($N=0$ and 1) \cite{Tscherbul:23}, and vibrational states \cite{Tesch:02, Zhao:06}.
	Robust QI storage is favoured by qubits that are not affected by the long-range ED interaction \cite{Park:17,Gregory:21}, whereas high-fidelity QI processing (such as robust two-qubit quantum gates) is easier to achieve with strongly interacting qubits \cite{Yelin:06,Ni:18}. The ability to switch between encodings yielding non-interacting and interacting qubits by means of, e.g., microwave pulses \cite{Ospelkaus:10a,Blackmore:20b} is a key component of QIS protocols based on ultracold molecules \cite{Ni:18}. This motivates the ongoing search of new qubit encoding schemes. 
At the same time, the discovery of new qubit encodings may give rise to novel applications of ultracold molecules for quantum computing, such as quantum annealing based on molecules \cite{Asnaashari:22}.

	%However, these are just two of the many possible encoding types available in molecules, with the vast majority of   options still unexplored. Examples include (i) non-adjacent rotational states such as $N=0$ and $N=2$,  (ii)   rotational states with $N$ different by more than 2, (iii) different Zeeman sublevels in the $N=0$ manifold of open-shell molecular radicals such as YO($^2\Sigma^+$) or CaF($^2\Sigma^+$),  etc. We will consider the first example in more detail in Section III.  Thus far, the potential of these (and other) new encodings has remained untapped.
	% and electronic states (Nature Phys). 
	%All of these encodings have their advantages and disadvantages. For example, rotational qubits are coupled by long-range ED interactions. Nuclear spin qubits have long coherence times, vibrational qubits have... RESUME HERE - THIS PARA NEEDS WORK
	%The great variety of possible qubit encodings realizable in molecules

	Despite the recent proposals \cite{Yelin:06,Gregory:21,Herrera:14,Karra:16,Tscherbul:23,Tesch:02, Zhao:06,Ni:18,Asnaashari:22}, there remains a wealth of possibilities for unexplored encodings  of qubits into molecular states. For example, qubits can be encoded into non-adjacent rotational levels (e.g., $N=0$ and 2) of the same or different vibronic state, into  hyperfine-Zeeman sublevels of the rovibronic states, or even into  different rovibrational and hyperfine states in different electronic manifolds.
	%can be used as the effective spin-1/2 system.
	%,  Zeeman  electron or nuclear spin projections, in different hyperfine states,? 
	This gives rise to the following question: What are the advantages and limitations of a given molecular qubit encoding for QI storage and processing?  This question is relevant because, as mentioned, QI storage and processing impose conflicting requirements on the type of encoding. 
	Specifically, good memory qubits must be well-isolated from an external environment, and their interactions with each other must be minimized to avoid decoherence \cite{Nielsen:10}. By contrast, good qubits for processing quantum information must be strongly interacting, with the  interactions controllable by external electromagnetic  fields \cite{Lemeshko:13}.

	Motivated by this question, we propose a classification of molecular qubit encodings based on the ED interaction between the qubits. We show that, given the matrix elements of the dipole moment operator in the single-molecule basis, it is possible to assess the potential utility of any given molecular qubit for  QI storage and processing applications.
	%:  (i) quantum information storage and (ii) engineering two-qubit quantum logic gates. 
	%The only information required for this assessment is that 
	As an example, in Sec. III, we  explore a new type of encoding of molecular qubit into non-adjacent rotational states ($N=0$ and 2), which gives rise to several types of encoding according to our classification scheme. 
	{In Section~IV, we show that similar arguments can be made for higher order long-range couplings by considering electric quadrupole interactions, which are essential for qubit encodings in nonpolar homonuclear molecules. }
	Section V concludes with a brief summary of our main results and discusses several directions for future work.

	\section{Classification of molecular qubit encodings}

	We consider an effective spin-1/2 system with the eigenstates $\ket{\downarrow}$ and $\ket{\uparrow}$  comprising a qubit. 	
%	Consider two different internal states of a polar diatomic molecule, $\ket{\downarrow}$ and $\ket{\uparrow}$, into which we intend to  encode an effective spin-1/2 system ($\ket{\downarrow}$ and $\ket{\uparrow}$ can be thought of as the logical states of a qubit).
	Choosing a particular encoding amounts to identifying $\ket{\downarrow}$ and $\ket{\uparrow}$  with  the physical  states of a diatomic molecule, such as electronic, vibrational, rotational, fine, hyperfine, Stark, or Zeeman states. The two-qubit, as well as many-qubit Hamiltonians inherit the properties of the molecular states used for the encoding.

	As the strong, anisotropic, and tunable electric dipole-dipole (ED) interaction between molecular qubits is central to their applications in QIS \cite{Wall:15c,Bohn:17}, we use the ED interaction as a basis of our classification. 
	The effective ED interaction between two isolated spin-1/2 systems encoded in molecules $i$ and $j$ takes the form of the XXZ Hamiltonian \cite{Gorshkov:11b,Wall:15c}
	%  in the two-qubit Hilbert space spanned by $\{\ket{\uparrow\uparrow}, \ket{\uparrow\downarrow}, \ket{\downarrow\uparrow}, 
	(see Appendix A for a derivation)
	\begin{equation}
		\hat{H}_{ij} = \frac{1-3\cos^2\theta_{ij}}{R^3} \left[\frac{J_\perp}{2}\left(\hat{S}_i^+ \hat{S}_j^- + \rm{h.c.}\right) + J_z \hat{S}_i^z \hat{S}_j^z \right. \left.+ W\left(\hat{\mathbb{I}}_i \hat{S}_j^z + \hat{S}_i^z \hat{\mathbb{I}}_j\right) + V\hat{\mathbb{I}}_i\hat{\mathbb{I}}_j\vphantom{\frac{J_\perp}{2}}\right],
		\label{eq:xxz_ham_main}
	\end{equation}
	where $\hat{S}_i^\alpha$ and $\hat{S}_j^\alpha$ are the effective spin operators acting in the two-dimensional Hilbert spaces of the $i$-th and $j$-th molecules ($\alpha=\pm, \, z$), $R_{ij}$ is the distance between the molecules and $\theta_{ij}$ is the angle between the quantization vector and the vector connecting the two molecules. 
	
	To derive Eq.~\eqref{eq:xxz_ham_main}, we assume that the ED interaction between molecules is much weaker than the energy difference between  $\ket{\uparrow}$ and $\ket{\downarrow}$. This is a realistic assumption for most molecules trapped in optical lattices and tweezers, for which $|R_{ij}|\geq 500$~nm, and the ED interaction rarely  exceeds 1~kHz. As a result, couplings that change the total angular momentum projection of two molecules give rise to energetically off-resonant transitions and can be neglected \cite{Wall:15c}. Examples of such processes include transitions, which  transfer angular momentum from  molecular rotation to their  relative (orbital) motion ($q=\pm1$ and $q=\pm 2$, see Appendix A). We further assume that the qubit states $\ket{\uparrow}$ and $\ket{\downarrow}$ are isolated, whether by symmetry or by energy detuning, from other molecular states. This is an essential requirement for any QIS protocol. 
	
	As shown in previous work (see, e.g., Refs. \cite{Gorshkov:11b,Wall:15c}) and detailed in Appendix A for the present discussion, the coupling constants in the effective spin-spin interaction Hamiltonian  \eqref{eq:xxz_ham_main} 
	\begin{align}
		J_z &= (d_\uparrow - d_\downarrow)^2, \label{eq:jz_main}\\
		J_\perp  &= 2d_{\uparrow\downarrow}^2 - |d^+_{\uparrow\downarrow}|^2 - |d^-_{\uparrow\downarrow}|^2, \label{eq:jperp_main}
	\end{align}
	can be expressed in terms of the matrix elements of the electric dipole moments (EDMs) of the individual molecules with the spherical tensor components $\hat{d}_p$ ($\hat{d}_0=\hat{d}_z$, $\hat{d}_{\pm1} = \mp(\hat{d}_x\pm i\hat{d}_y)/\sqrt{2}$):
	\begin{align}
		\braket{\uparrow|\hat{d}_0|\uparrow} &\equiv d_\uparrow, \quad\,\,\,\,\,\,\,\,\, \braket{\downarrow|\hat{d}_0|\downarrow} \equiv d_\downarrow, \label{eq:dipoles1}\\
		\braket{\uparrow|\hat{d}_0|\downarrow}&\equiv d_{\uparrow\downarrow}, \quad \braket{\uparrow|\hat{d}_{\pm 1}|\downarrow}\equiv d^\pm_{\uparrow\downarrow}.\label{eq:dipoles2}
	\end{align}
	Significantly, the Ising coupling constant $J_z$ depends on the {\it difference between the diagonal matrix elements of the EDM} in the qubit basis, and the spin exchange coupling $J_\perp$ scales with the square of the {\it off-diagonal (or transition) matrix element}.
	% of the EDM between the qubit states.
	In practice, in the absence of mixing between angular momentum projection states, either $d_{\uparrow\downarrow} $ or  $d^\pm_{\uparrow\downarrow}$ vanish, so either  the first or  the last two terms on the right-hand side of Eq.~\eqref{eq:jperp_main} are different from zero.
	The terms parameterized by the constants $W = (d_\uparrow^2 - d_\downarrow^2)/2$ and $V= (d_\uparrow + d_\downarrow)^2/4$ in Eq.~\eqref{eq:xxz_ham_main} result in the overall energy shift for a homogeneous ensemble of pinned molecules \cite{Wall:15c}, so we neglect them in the following.
	
	The effective ED interaction in the form of Eq.~(\ref{eq:xxz_ham_main}) can be used in combination with Eqs.~\eqref{eq:jz_main}-\eqref{eq:jperp_main} to classify the different qubit encodings. To this end, we first note that if $J_z=0$ and $J_\perp=0$, the ED interaction between  the qubits is identically zero. According to Eq.~\eqref{eq:xxz_ham_main} the vanishing of the ED interaction requires the following two conditions to be simultaneously fulfilled: $d_\uparrow = d_\downarrow$ and $d_{\uparrow\downarrow}= 0$. Thus,  all qubit encodings,  for which the diagonal elements of the EDM are equal and the off-diagonal matrix elements  vanish,  will have zero ED interaction. Because of this, we expect such qubit encodings  to have long coherence times, which can be advantageous for  long-term quantum information storage applications  (memory qubits). 
	%Therefore, all qubit encodings, for which $d_\uparrow = d_\downarrow$ and $d_{\uparrow\downarrow}= 0$
	% label each encoding with
	
	We use a pair of categorical variables $Z$ and $X$ to characterize the encodings based on Eq.~(\ref{eq:xxz_ham_main}).  The variable $Z$ takes the value of 0 if the Ising interaction is zero  ($J_z=0$) and 1 otherwise. Similarly, the variable $X$ takes the value of 0 if the spin-exchange interaction is zero   ($J_\perp=0$) and 1  otherwise.  This gives rise to four possible encodings listed in Table 1.
	For example, the  encoding, for which $J_z=J_\perp=0$, is classified as 0/0. We can also refer to it as    {\it interactionless}  because, as shown above, the qubit states  are not coupled by the long-range ED interaction.

	As an example of the 0/0 encoding, consider the nuclear spin sublevels of the ground ($N=0$) rotational state of alkali-dimer molecules. In the high magnetic field limit, the eigenstates of these molecules can be written as $|NM_N\rangle|I_1M_{I_1}\rangle | I_2M_{I_2}\rangle = |NM_N, M\rangle$, 
	where $|NM_N, M\rangle$ are the eigenstates of rotational angular momentum $\hat N$ and its $z$-component $\hat N_z$, 
	 $|I_\alpha M_{I_\alpha}\rangle$ are the  eigenstates of the nuclear spin operators $\hat{I}_\alpha^2$ and $\hat{I}_{\alpha_z}$ of the $\alpha$-th nucleus ($\alpha=1,2$), and $M=\{M_{I_1}, M_{I_2}\}$ is a collective nuclear spin quantum number \cite{Aldegunde:08}. The nuclear spin qubit states are then encoded as $\ket{\uparrow}=\ket{00,M}$ and $\ket{\downarrow}=\ket{00,M'}$ with $M\ne M'$. With this encoding,  $d_{\uparrow}=d_{\downarrow}$  because both the $\ket{\uparrow}$ and $\ket{\downarrow}$ states have $N=0$, and $d_{\uparrow\downarrow}=0$ because the EDM operators  $\hat{d}$ and $\hat{d}_\pm$ are diagonal in the nuclear spin quantum number (for concreteness, we focus on the $q=0$ spherical tensor component of the EDM operator $\hat{d}=\hat{d}_0$, noting that the $q={\pm 1}$ components can be treated in a similar way)
	\begin{equation}\label{EDM_mx_el}
		\langle NM_N, M | \hat{d} | N'M_N', M'\rangle = \langle NM_N | \hat{d} | N'M_N'\rangle \delta_{MM'}
	\end{equation}
	and the nuclear spin qubit states have $M\ne M'$. Here, $\delta_{MM'}=\delta_{M_{I_1}M_{I_{1}}'}\delta_{M_{I_2}M_{I_{2}}'}$.
	In the presence of an external dc electric field $E$, the expectation values $d_{\uparrow}$ and  $d_{\downarrow} $ are different from zero, but  $d_{\uparrow}=d_{\downarrow}$, so both $J_z$ and $J_\perp$ vanish even at $E>0$. This effectively cancels  the long-range ED interactions, leading to long coherence times of several seconds or longer, as observed experimentally for ultracold trapped KRb \cite{Ospelkaus:10a,Yan:13}, RbCs \cite{Gregory:21}, NaK\cite{Park:17}, and NaRb \cite{Lin:22} molecules.

	As noted above, in order for two qubits to interact via the long-range ED interaction \eqref{eq:xxz_ham_main},  either $J_z$  or $J_\perp$ (or both) must be nonzero.  This leads to three other types of encodings listed in Table 1, which we now proceed to analyze.
	
	First, if  $J_z=0$ and $J_\perp \neq 0$,  the effective spin-spin interaction between the $i$-th and $j$-th qubits \eqref{eq:xxz_ham_main}  takes the  form of the long-range spin-exchange interaction \cite{Gorshkov:11a,Gorshkov:11b,Wall:15c}
	\begin{equation}
		H_{ij} = \frac{1-3\cos^2\theta_{ij}}{R^3} \left[\frac{J_\perp}{2}\left(S_i^+S_j^- + \rm{h.c.} \right)\right],
		\label{eq:xxz_ham_spin_ex}
	\end{equation}
	Because  $J_\perp\ne 0$ and $J_z=0$, this encoding can be classified as 0/1 (see Table 1).  This is by far the most common type of encoding considered in the literature to date. As an example, the encoding into adjacent rotational states  $\ket{\uparrow}=\ket{00,M}$ and $\ket{\downarrow}=\ket{1M_N,M'}$ with $M= M'$  was originally proposed in a seminal paper by DeMille \cite{DeMille:02}. 
	%This choice of encoding corresponds to.
	 It follows from Eq.~\eqref{EDM_mx_el} that the off-diagonal EDM matrix element does not vanish
	\begin{equation}
		d_{\uparrow\downarrow} = \langle 00, M | \hat{d} | 1 M_N, M\rangle = \langle 00 | \hat{d} | 1 M_N\rangle \ne 0,
	\end{equation}
	and hence  $J_\perp \propto d_{\uparrow\downarrow}^2\ne 0$. The diagonal EDM matrix elements in the adjacent rotational state encoding are zero in the absence of an external $E$ field because the two rotational states have a definite parity. As a result, $J_z= 0$ and the adjacent rotational state encoding can be classified as 0/1.
	In nonzero electric field, $d_{\uparrow}\ne d_{\downarrow}$ and thus $J_z\ne 0$, and the type of encoding changes to 1/1 (see below). This demonstrates that applying an external $E$ field can cause interconversion between the different  types of encodings. We will see another example of this ``encoding crossover'' in Sec. III.
	% can interconvert depending on whether an external field is present.

	Second, if  $J_z\ne0$ and $J_\perp= 0$,  the effective spin-spin interaction between the $i$-th and $j$-th qubits \eqref{eq:xxz_ham_main}  takes the  form  of  the long-range Ising interaction \cite{Tscherbul:23}
	\begin{equation}
		H_{ij} = \frac{1-3\cos^2\theta_{ij}}{R^3}  J_zS_i^zS_j^z,
		\label{eq:Ising_ham_main}
	\end{equation}
	Because $J_z\ne0$ and $J_\perp= 0$, this is a 1/0 encoding (see Table 1). Until very recently, this type of encoding has been virtually unexplored, unlike the  standard 0/0 and 0/1  encodings  \cite{Gregory:21,DeMille:02}.
	% into the nuclear spin and (especially the adjacent rotational state encoding). 
	One example of such encoding, which we will refer to as spin-rotational, can be realized by the lowest two rotational states with {\it different} nuclear spin projections \cite{Tscherbul:23}, i.e., $\ket{\uparrow}=\ket{00,M}$ and $\ket{\downarrow}=\ket{1M_N,M'}$ with $M \ne M'$ (note that in the standard encoding into adjacent rotational states, $M=M'$).
	
	Equation \eqref{eq:Ising_ham_main} shows that encodings of the 1/0 type, such as the  spin-rotational encoding, naturally give  rise to the long-range Ising interaction \cite{Tscherbul:23}. Dynamical evolution of quantum many-body systems interacting via the Ising Hamiltonian  generates cluster-state entanglement \cite{Briegel:01}. Cluster states are universal entangled resource states for measurement-based quantum computation \cite{Briegel:01,Briegel:09}. The Ising interaction \eqref{eq:Ising_ham_main} can also be used to implement universal two-qubit  quantum logic gates (Ising gates), which  have been explored in the context of  nuclear magnetic resonance (NMR)-based quantum computation \cite{Jones:02}. These interactions can also be used for more complex QI protocols, such as, for example, protocols based on qubits encoded into states of multiple molecules, which can be used for engineering transverse-field Ising models for applications such as quantum annealing \cite{Asnaashari:22}. 
	
	Finally, when both $J_\perp$ and $J_z$ are nonzero, the effective spin-spin interaction contains both the Ising and spin-exchange terms, leading to the 1/1 type encoding, in which all the terms in the XXZ Hamiltonian are nonzero. As stated above, one common example of such  encoding is furnished  by  the adjacent rotational states with the same nuclear spin projection ($\ket{\uparrow}=\ket{00,M}$ and $\ket{\downarrow}=\ket{1M_N,M'}$ with $M = M'$) in nonzero electric field (in Sec. III, we will consider a less familiar example of encoding into non-adjacent rotational states). In this encoding, $d_{\uparrow\downarrow}\ne 0$ and the presence of the field ensures that $d_{\uparrow}\ne d_{\uparrow}$, and thus $J_z\ne 0$. The property of both the Ising and spin-exchange interactions being different from zero is advantageous for a number of applications, such as dynamical generation of many-body spin-squeezed states \cite{Bilitewski:21,Tscherbul:23}, which can be used to achieve metrological gain over the standard quantum limit \cite{Ma:11,Pezze:18}. At the same time, the coherence properties of 1/1 type qubits (as well as those of the 1/0 and 0/1 types) may be limited by the strong and long-range ED interaction, which is experimentally challenging to turn on and off.
	
	% for measurements beyiong  

%	Table 1 summarizes the different classes of qubit encodings defined above, and lists their potential applications.

	\begin{table}[t]
		\begin{center}
			\begin{tabular}{cccc}
				\hline \hline
				Qubit type & Examples & Advantages & Limitations  \\
				\hline
				0/0 (interactionless) & Nuclear spin sublevels ($N = 0$) & Coherence time & Two-qubit gates \\    
				$d_\uparrow=d_\downarrow$, $d_{\uparrow\downarrow}=0$  & Non-adjacent rotational states, $E=0$ &  QI storage  &  \\  
				\hline
				0/1 (spin-exchange) & Adjacent rotational states, $E=0$$^{(1)}$ & Two-qubit gates & Coherence time \\    
				$d_\uparrow=d_\downarrow$, {$d_{\uparrow\downarrow}\ne 0$} &  & Entanglement   &   \\  
				\hline
				1/0 (Ising) 	       & Spin-rotational states ($N = 0,1$)$^{(2)}$  & Two-qubit gates & Coherence time \\    
				{$d_\uparrow\ne d_\downarrow$}, {$d_{\uparrow\downarrow} = 0$} & Non-adjacent rotational states$^{(3)}$ & Entanglement   &   \\  			       
				\hline
				1/1 (XXZ) 	               & Adjacent rotational states, $E>0$$^{(1)}$  & Two-qubit gates & Coherence time \\    
				{$d_\uparrow\ne d_\downarrow$, {$d_{\uparrow\downarrow} \ne 0$}} & Non-adjacent rotational states, $E>0$$^{(4)}$ & Entanglement   &   \\ 
				\hline \hline
			\end{tabular}
			\caption{Qubit encoding classification based on the effective ED interqubit coupling.}
		\end{center}
		\label{tab:Encoding_classes}
		\vspace{-7mm}
		\flushleft$^{(1)}$With the same spin projection ($M=M'$).
		\vspace{-3mm}
		\flushleft$^{(2)}$With different  spin projections ($M\ne M'$).
		\vspace{-3mm}
		\flushleft$^{(3)}$With $|\Delta M_N | \ge 2$ regardless of $M$ and $M'$.
		\vspace{-3mm}
		\flushleft$^{(4)}$With $|\Delta M_N |\leq 1$ and the same spin projection ($M=M'$) .
		
	\end{table}
	
	%If only  $J_\perp\ne 0$ then we get spin-exchange interactions, spin-exchange encoding (0/1).
	%If both $J_z\ne 0$  and $J_\perp\ne 0$, we get the full XXZ  encoding (1/1).
	
	% These conditions are met if  $d_\uparrow\ne d_\downarrow$ (which implies $J_z\ne 0$) or $d_{\uparrow\downarrow}\ne 0$ (which implies $J_\perp\ne 0$). Each of these two situations leads to a distinct type of ED interaction, i.e., either, Ising  spin-exchange, or both. 
	
	%The Ising interaction (Z) is present if $d_\uparrow\ne d_\downarrow$. We can therefore label each qubit encoding with a categorical variable (Z), which is 0 if $J_z=0$, and 1 if $J_z>0$.
	
	%The spin-exchange interaction (X) is present if $d_{\uparrow\downarrow}\ne 0$.  \textcolor{red}{(This needs to be updated based on the general result for $J_\perp$ from the previous section).} A qubit econding can therefore be assigned another categorical variable X, which is 0 if $J_\perp=0$, and 1 if $J_\perp \ne 0$.
	
	%Based on the values of the categorical variables Z and X, we can define 4 possible types of qubit encodings (Z/X):  0/0, 0/1, 1/0, and 1/1, listed in Table I.
	%Finally, we observe that while all four terms depend on the matrix elements of the $\hat{d}_0$ operator, $J_\perp$ depends on {\it both} the transition dipole matrix elements of $\hat{d}_0$ and $\hat{d}_{\pm 1}$. 

	\section{Qubit encoding into non-adjacent rotational levels}
	
\begin{figure}[t]
		\includegraphics[width=0.8\columnwidth]{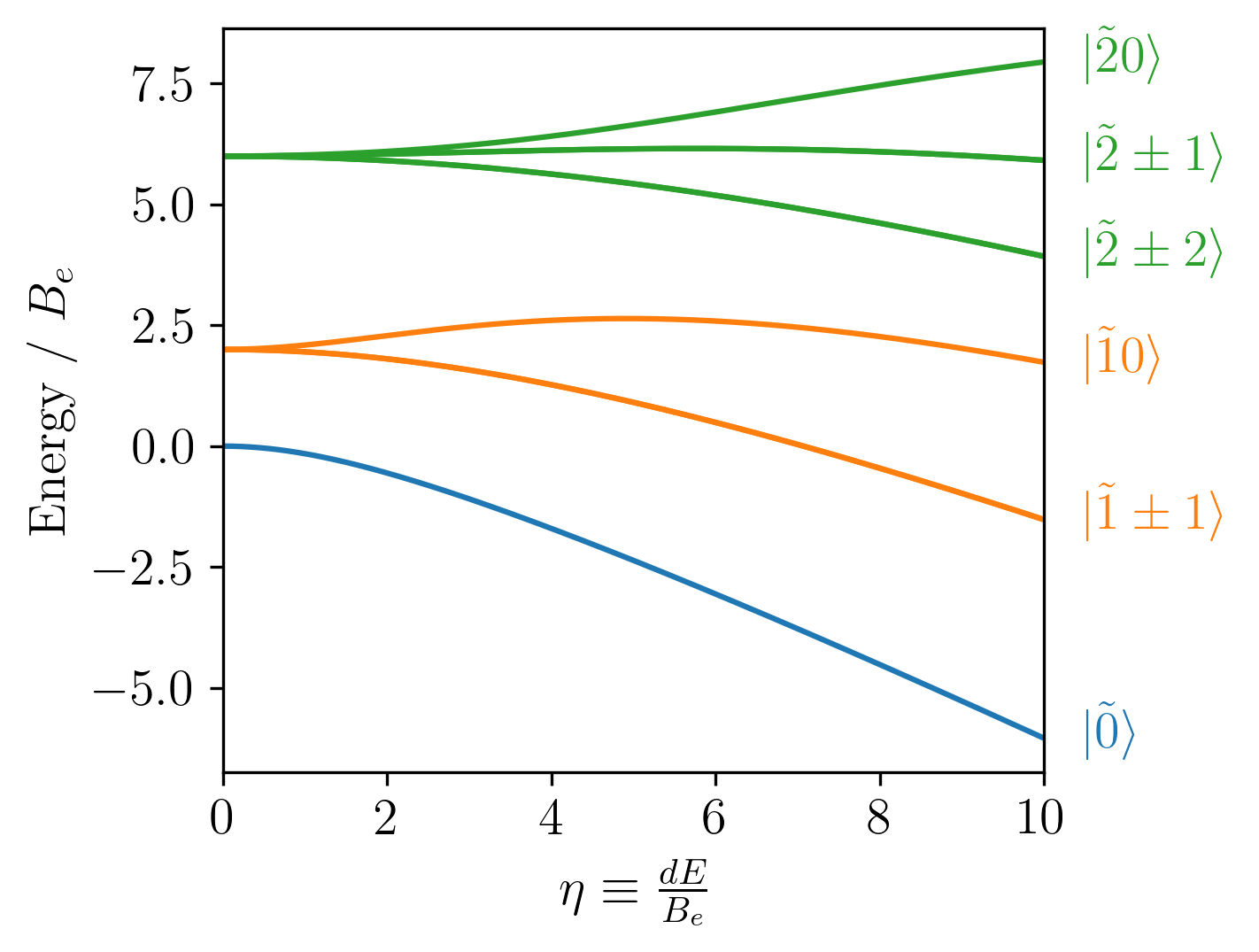}
		\caption{Energy levels of a ${}^1\Sigma$ molecule as functions of the effective electric field.}
		\label{fig:energy_levels}
	\end{figure}

	To illustrate the application of the proposed classification scheme, we consider a previously unexplored type of qubit encoding  into non-adjacent rotational levels of a ${}^1\Sigma$ molecule. The Hamiltonian of a ${}^1\Sigma$ molecule in the vibrational ground state placed in dc electric field $\bm E$ can be written as
	\begin{equation}
		\hat{H} = B_e \hat N^2 - \bm E\cdot\bm d\label{eq:1s_ham}
	\end{equation}
	where $B_e$ is the rotational constant, and $\bm d$ is the dipole moment of the molecule. Figure \ref{fig:energy_levels} shows the lowest nine eigenvalues of Hamiltonian \eqref{eq:1s_ham} as functions of the effective electric field $\eta\equiv {dE}/{B_e}$. These states correspond to rotational states with $N=0, 1$ and $ 2$ at zero field.
	
	The interaction of molecules with the electric field couples rotational states $\ket{NM_N}$ with the same projection ($M_N$) to yield dc field-dressed (or pendular) states   \cite{Rost:92,Wall:15c}
	\begin{equation}
		\ket{\tilde{N}M_N} = \sum_{N} c_{N}\ket{N M_N}.
	\end{equation}
	
		\begin{figure}[t]
		\includegraphics[width=0.8\columnwidth]{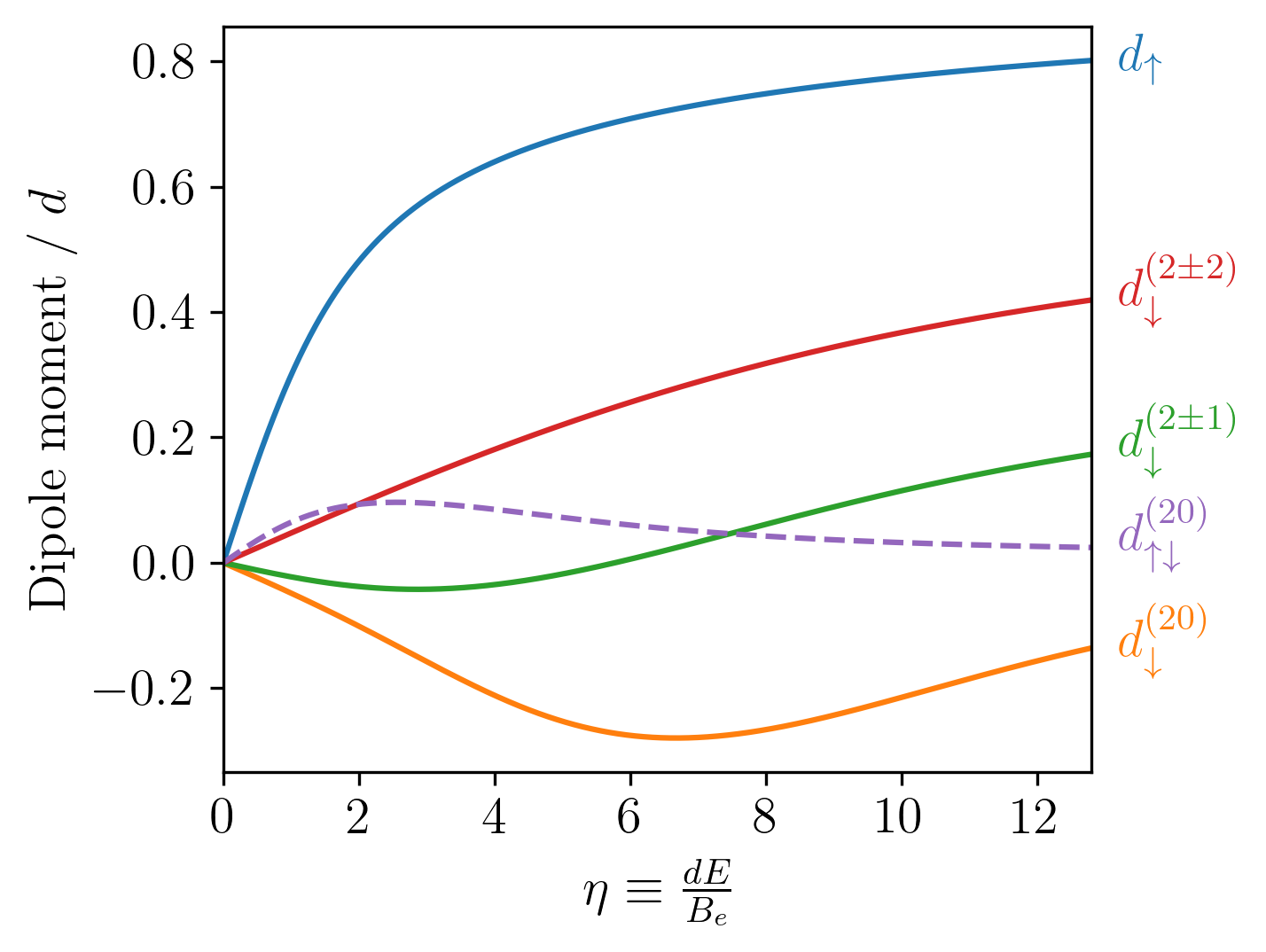}
		\caption{Non-zero EDM matrix elements of $\hat{d}_0$ for a $^1\Sigma$ molecule as a function of the effective electric field.}
		\label{fig:dipoles}
	\end{figure}
	
	The choice of $\ket{\downarrow} = \ket{\tilde{0}0}$ leaves five choices for $\ket{\uparrow}$ from the manifold correlating at zero field with $N=2$, including $\ket{\tilde{2}0}, \ket{\tilde{2}\pm1}$ and $\ket{\tilde{2}\pm2}$. The matrix elements of the dipole operators in the rotational basis states are calculated as
	\begin{equation}
		\braket{N'M_{N'}|\hat{d}_p|NM_N} = d(-1)^{M_{N'}}\sqrt{(2N'+1)(2N+1)} 
		\begin{pmatrix}
			N' & 1 & N \\
			-M_{N'} & p & M_N
		\end{pmatrix}\begin{pmatrix}
			N' & 1 & N \\
			0 & 0 & 0
		\end{pmatrix}\label{eq:dipole_elements}
	\end{equation}
	where $d$ is the permanent dipole moment of the molecule. The $3j$-symbols in Eq. \eqref{eq:dipole_elements} vanish if $|N' - N| < 1$ and $M_N + p \neq M_{N'}$. 
	The non-zero EDM matrix elements of $\hat{d}_0$ as a function of the effective electric field $\eta\equiv{dE}/{B_e}$ are displayed for the three possible encodings in Fig. \ref{fig:dipoles}, where 
	\begin{align}
		d_\uparrow &= \braket{\tilde{0}0|\hat{d}_0|\tilde{0}0}, \\
		d_\downarrow^{(NM_N)} &= \braket{\tilde{N}M_N|\hat{d}_0|\tilde{N}M_N}, \\
		d_{\uparrow\downarrow}^{(NM_N)} &= \braket{\tilde{0}0|\hat{d}_0|\tilde{N}M_N}.
	\end{align}
The transition matrix elements of $\hat{d}_p$ ($d_{\uparrow\downarrow}$ and $d_{\uparrow\downarrow}^\pm$) vanish for states with $M_N\neq p$. Therefore, $d_{\uparrow\downarrow}$ is only non-zero when $\ket{\downarrow} = \ket{\tilde{2}0}$ and $d_{\uparrow\downarrow}^\pm$ is only non-zero when $\ket{\downarrow} = \ket{\tilde{2}\pm1}$. 
Note that $d_{\uparrow\downarrow}\ne 0$ at $E>0$ because the  states $\ket{\tilde{0}0}$ and $\ket{\tilde{2}0}$ couple through the intermediate state $\ket{\tilde{1}0}$. 
All transition dipole matrix elements vanish for $\ket{\downarrow} = \ket{\tilde{2}\pm2}$. All diagonal matrix elements are zero for bare rotational states at zero field.

		\begin{figure*}[t]
		\includegraphics[width=\textwidth]{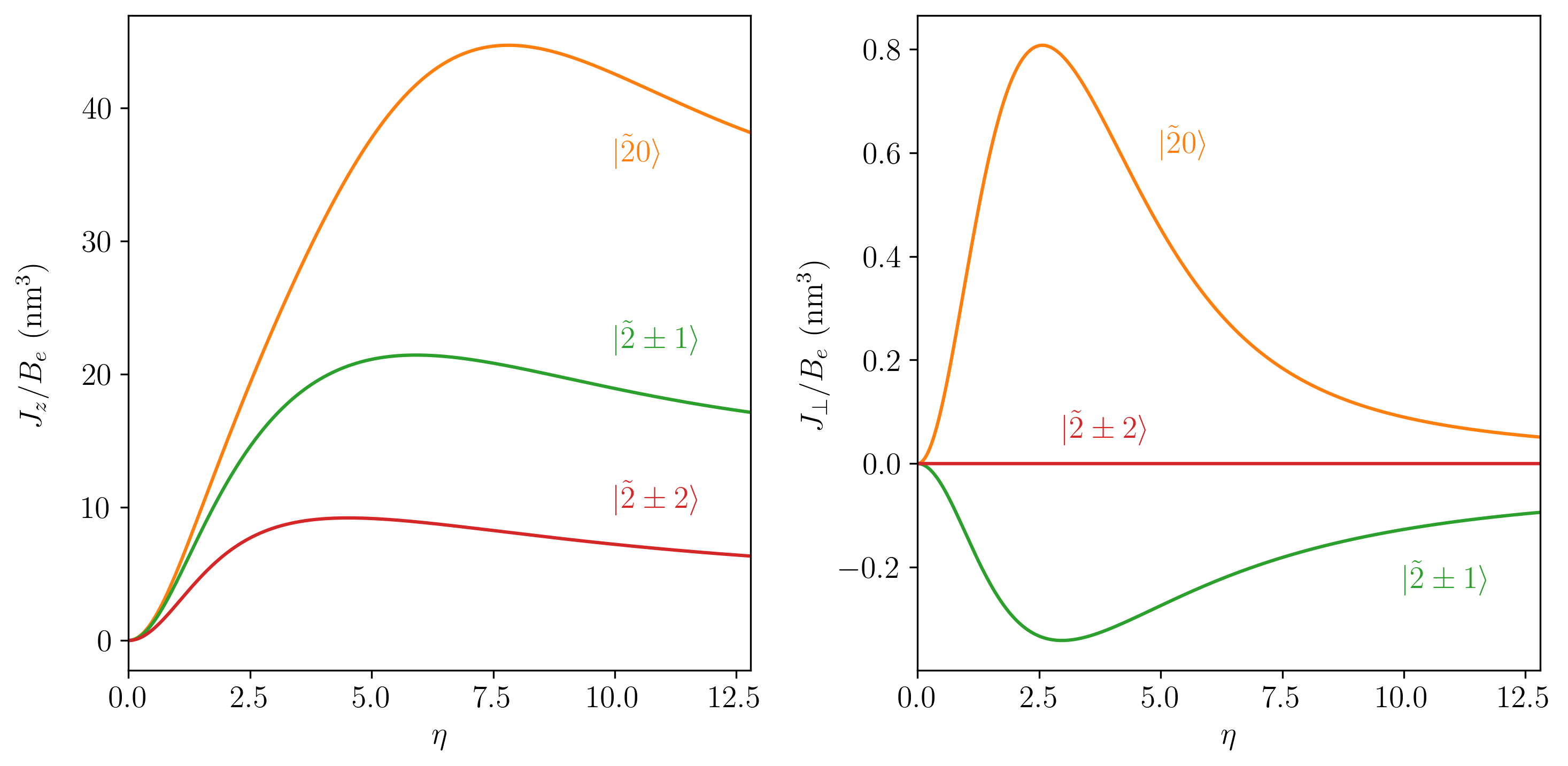}
		\caption{Couplings $J_z$ (left) and $J_\perp$ (right) as a function of the effective electric field, using the ground rotational state and the indicated rotational states of a polar $^1\Sigma$ molecule to encode the qubits.}
		\label{fig:couplings}
	\end{figure*}
	
	As follows from Fig. \ref{fig:dipoles}, qubits spanning the $\tilde N = 0, \tilde N = 2$ pairs permit two classes of encoding: the 0/0 encoding at vanishing electric field, 1/0 encoding and 1/1 encoding, depending on the $\ket{\uparrow}$  state. 
	%{\color{red} Please check/extend the preceding sentence - it kind of gives a special meaning to teh $N=0,2$ encoding in this paper.}
	To illustrate this more clearly, we calculate the parameters of the XXZ interaction of Eq.~\eqref{eq:xxz_ham_main} using   Eqs.~(\ref{eq:jz_main})-(\ref{eq:jperp_main}) and the EDM matrix elements. Figure \ref{fig:couplings} shows the couplings $J_z$ and $J_\perp$ as functions of the effective electric field. As described in Eq. \eqref{eq:jz_main}, the Ising coupling between qubits grows as a function of the difference in the diagonal matrix elements of the qubit states. This is observed in the left panel of Fig. \ref{fig:couplings}, as $J_z$ is consistently maximized for $\ket{\downarrow} = \ket{\tilde{2}0}$. On the other hand, $J_\perp$ corresponds to the square of the off-diagonal EDM matrix elements. As expected, $J_\perp$ is equal to zero when $\ket{\downarrow} = \ket{\tilde{2}2}$ regardless of the $E$-field, while $J_\perp$ is positive for $\ket{\downarrow} = \ket{\tilde{2}0}$ and negative for $\ket{\downarrow} = \ket{\tilde{2}0}$ at non-zero $E$-field. However, both perpendicular couplings are maximized at intermediate field strengths, and diminish in strong fields.

\section{Quadrupole-quadrupole interaction}
	
	The classification scheme proposed here can be extended to other long-range interactions than the ED interaction. As an example, we consider in the present section the quadrupole-quadrupole (QQ) interaction. This interaction is the leading long-range interaction between homonuclear molecules,  
which do simultaneously possess even/odd $N$ state manifolds and which therefore can particularly benefit from the qubit encoding introduced in the preceding section. Previous theoretical work has shown that QQ interactions of nonpolar atoms and molecules in two-dimensional optical lattices can give rise to exotic topological phases \cite{Bhongale:13}.

As shown in Appendix B, the QQ interaction leads to the same XXZ model as given by Eq.~(\ref{eq:xxz_ham_main}). However, the model parameters must now be expressed in terms of the matrix elements of the quadrupole moment, yielding the following relations:
	\begin{align}
		J_z &= (q_\uparrow - q_\downarrow)^2, \label{eq:jz_q}\\
		J_\perp &= 2\left[6q_{\uparrow\downarrow}^2 - 4 \left[|q_{\uparrow\downarrow}^{+1}|^2 + |q_{\uparrow\downarrow}^{-1}|^2\right] + \left[|q_{\uparrow\downarrow}^{+2}|^2 + |q_{\uparrow\downarrow}^{-2}|^2\right]\right], \label{eq:jperp_q}
	\end{align}
Thus, the same classification scheme can be applied to qubits encoded in non-polar molecules interacting through QQ couplings with analogous conditions on the vanishing of $J_z$ and $J_\perp$, which can be analyzed by considering the matrix elements of the quadrupole moment in the single-molecule basis.

	\begin{figure}[t!]
		\includegraphics[width=0.8\columnwidth]{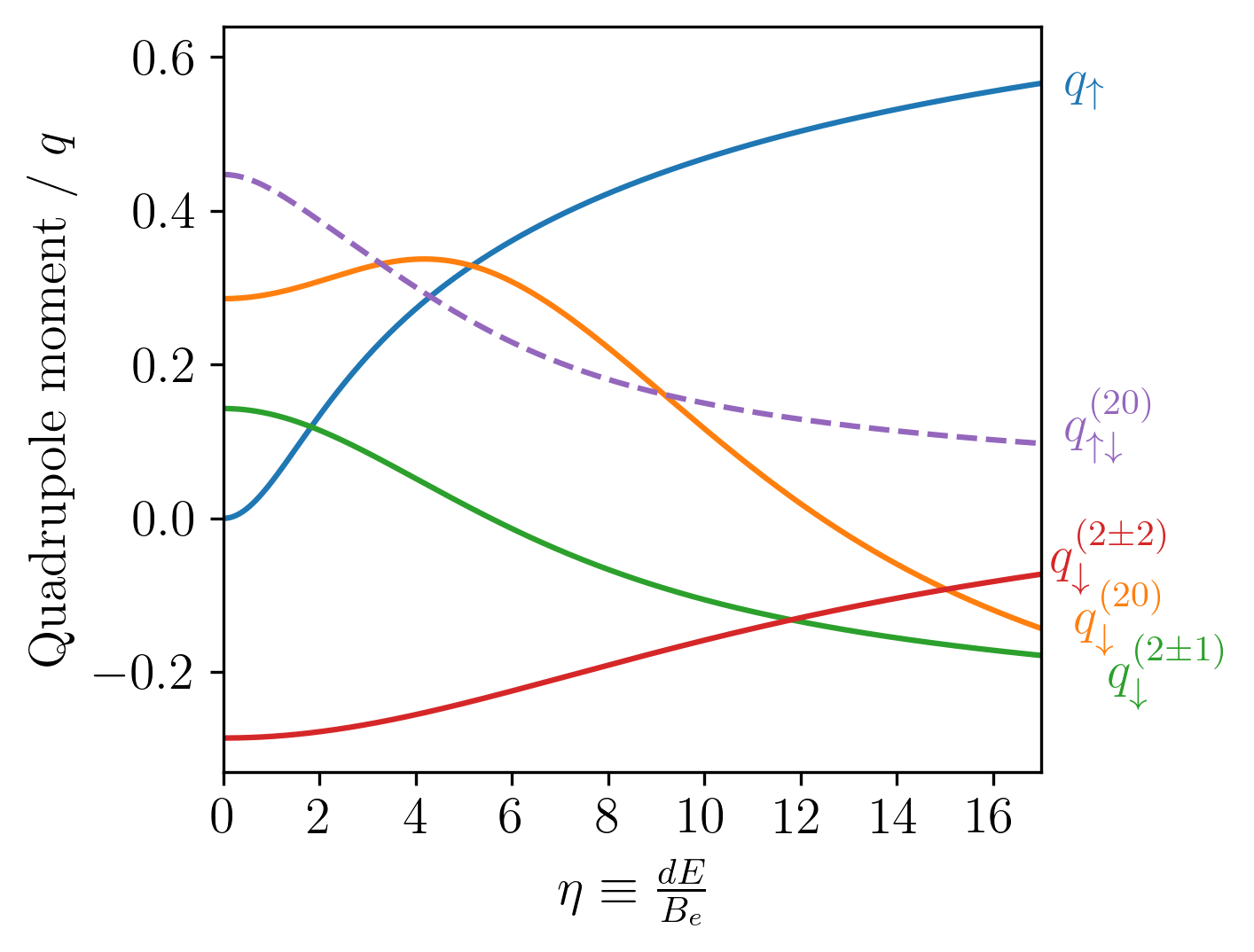}
		\caption{Non-zero quadrupole matrix elements of $\hat{q}_0$ as a function of the effective electric field.}
		\label{fig:quadrupoles}
	\end{figure}

The matrix elements of the quadrupole operators in the rotational basis states are calculated as
	\begin{align}
		\braket{N'M_{N'}|\hat{q}_p|NM_N} =& q(-1)^{M_{N'}}\sqrt{(2N'+1)(2N+1)} \nonumber \\
		&\times\begin{pmatrix}
			N' & 2 & N \\
			-M_{N'} & p & M_N
		\end{pmatrix}\begin{pmatrix}
			N' & 2 & N \\
			0 & 0 & 0
		\end{pmatrix}\label{eq:quadrupole_elements}
	\end{align}
	where $q$ is the permanent quadrupole moment of the molecule. The $3j$-symbols in Eq. \eqref{eq:quadrupole_elements} vanish if $|N' - N| < 2$ and $M_N + p \neq M_{N'}$. 
	The non-zero EDM matrix elements of $\hat{q}_0$ as a function of the effective electric field $\eta\equiv\frac{dE}{B_e}$ are displayed for the three possible encodings in Fig. \ref{fig:quadrupoles}, where 
	\begin{align}
		q_\uparrow &= \braket{\tilde{0}0|\hat{q}_0|\tilde{0}0}, \\
		q_\downarrow^{(NM_N)} &= \braket{\tilde{N}M_N|\hat{q}_0|\tilde{N}M_N}, \\
		q_{\uparrow\downarrow}^{(NM_N)} &= \braket{\tilde{0}0|\hat{q}_0|\tilde{N}M_N}.
	\end{align}
The transition matrix elements of $\hat{q}_p$ ($q_{\uparrow\downarrow}$, $q_{\uparrow\downarrow}^{\pm1}$ and $q_{\uparrow\downarrow}^{\pm2}$) vanish for states with $M_N\neq p$. Therefore, $q_{\uparrow\downarrow}$ is only non-zero when $\ket{\downarrow} = \ket{\tilde{2}0}$, $q_{\uparrow\downarrow}^{\pm1}$ is only non-zero when $\ket{\downarrow} = \ket{\tilde{2}\pm1}$ and $q_{\uparrow\downarrow}^{\pm2}$ is only non-zero when $\ket{\downarrow} = \ket{\tilde{2}\pm2}$.

	\begin{figure*}[t]
		\includegraphics[width=\textwidth]{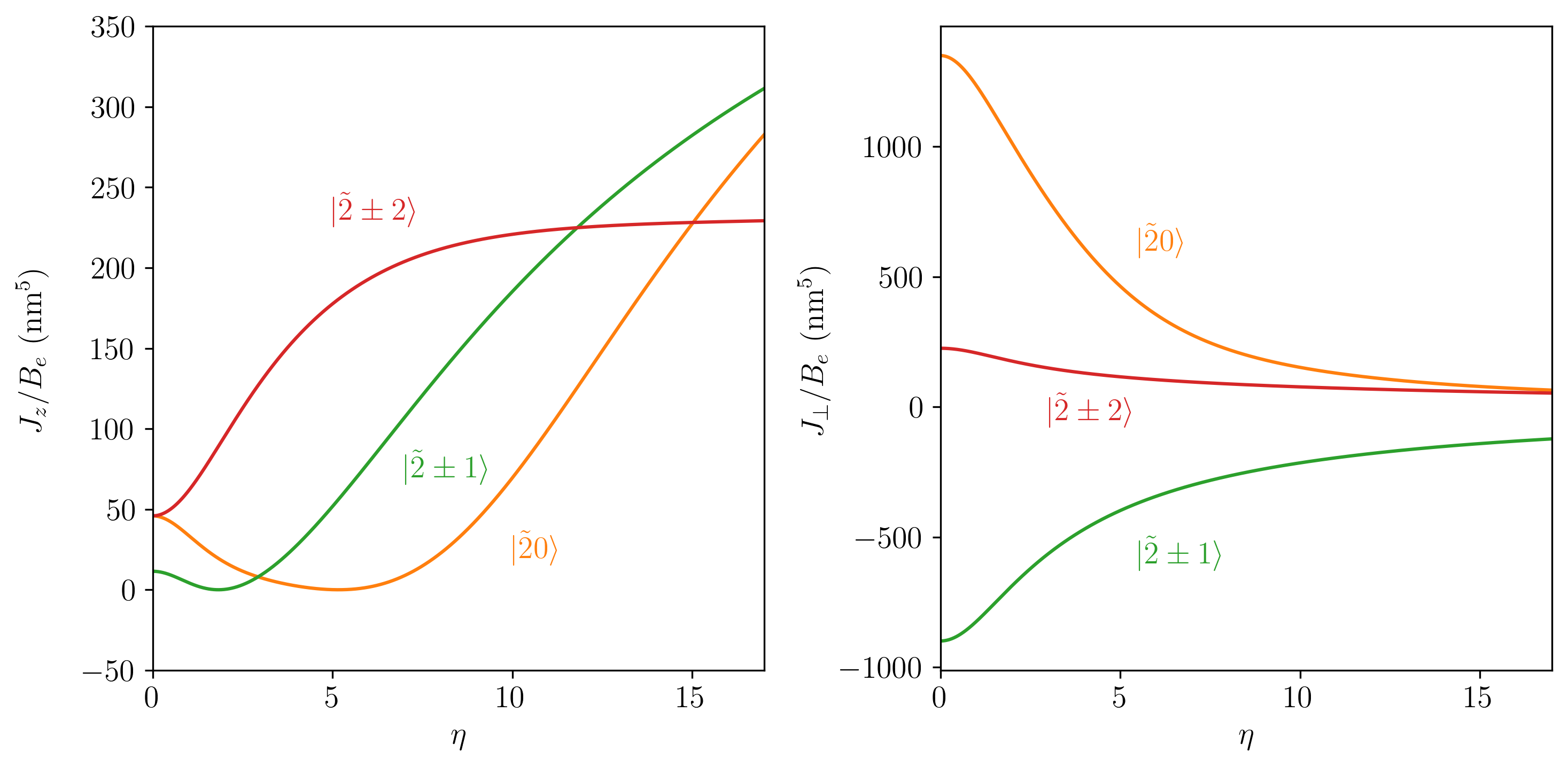}
		\caption{Quadrupole contribution to couplings $J_z$ (left) and $J_\perp$ (right) as a function of the effective electric field, using the ground rotational state and the indicated rotational states to encode the qubits.}
		\label{fig:couplings_q}
	\end{figure*}

 To illustrate the possible types of encoding achievable with homonuclear molecules, %as described in Section III, 
 we calculate the parameters of the XXZ interaction of Eq. \eqref{eq:xxz_ham_main} using Eqs. (\ref{eq:jz_q}-\ref{eq:jperp_q}) and the quadrupole matrix elements. Figure \ref{fig:couplings_q} shows the couplings $J_z$ and $J_\perp$ as functions of the effective electric field. As described in Eq. \eqref{eq:jz_q}, the Ising coupling between qubits grows as a function of the difference in the diagonal matrix elements of the qubit states. On the other hand, $J_\perp$ corresponds to the square of the off-diagonal quadrupole matrix elements. As expected, $J_\perp$ is positive for $\ket{\downarrow} = \ket{\tilde{2}0}$ and $\ket{\downarrow} = \ket{\tilde{2}\pm2}$ and negative for $\ket{\downarrow} = \ket{\tilde{2}\pm1}$ at non-zero $E$-field. However, all three perpendicular couplings are maximized at low field strengths, and diminish in strong fields. It can be observed that the QQ interaction permits the following types of encoding based on the $\tilde N = 0$ and $\tilde N = 2$ states: 0/1, 1/1 and 1/0. 
 
 Finally, we note that for typical lattice spacings used in current  experiments ($\simeq 500-1000$~nm), the QQ interaction is on the order of $\leq$1 Hz, which is  two-three orders of magnitude weaker than the ED interaction between polar molecules. 
  Despite its weakness, the effective QQ interaction Hamiltonian has the same form as the ED Hamiltonian, 
and can therefore be used (at least in principle)  to  generate useful many-body entangled states of nonpolar molecules in the same way as the ED interaction of polar molecules \cite{Bilitewski:21,Tscherbul:23}. In order to ensure robust dynamical evolution toward such entangled states,  the evolution timescale should be much shorter than the coherence time of non-adjacent rotational state superpositions \cite{Micheli:06,Tscherbul:23}.
% in a coherent superposition for a sufficiently long time.

 % than 100 times  smaler than the 
 %{\color{red} Please check this last sentence}. 

\section{Conclusions}

Molecules are complex quantum systems that feature multiple energy scales ranging from tens of Hz (hyperfine, Zeeman, and tunneling doublet structure) to thousands of THz (electronic structure). In addition, intermolecular interactions at short range are described by multidimensional potential energy surfaces (PESs), whose accurate description requires sophisticated quantum chemistry techniques and fitting methods.
However, the long-range physics of intermolecular interactions of relevance to current QIS  experiments in optical lattices and tweezers \cite{Bohn:17,Kaufman:21} is completely described by the well-established multipole  expansion. The lowest leading order in the multipole expansion for neutral (uncharged) polar molecules  is represented by  the ED  interaction, and  the next leading orders by the EQ and QQ interactions.

For qubit-based QIS applications, the molecule is reduced to a two-level system (the qubit), whose effective spin-1/2 levels can be encoded into the electronic, vibrational, rotational, fine, hyperfine, Stark, or Zeeman states.  Different choices of encoding give rise to different flavors of the ED interaction between the qubits. 
 Here, we have shown that  all possible encodings can be classified into 4 types based on the flavor of the effective ED interaction they give rise to.

 Our classification is based on two realizations. First, the general interaction between molecular qubits is completely determined   by the ED Hamiltonian in the effective two-qubit basis. Second, the form of this Hamiltonian depends only on the matrix elements of the   EDM of the individual molecules in a given encoding. The general ED Hamiltonian takes the form of the iconic XXZ  Heisenberg model of quantum magnetism \cite{Wall:15c} with the Ising and spin-exchange coupling parameters  $J_z$  and  $J_\perp$  expressed in terms of the diagonal ($d_\uparrow$, $d_\downarrow$) and off-diagonal ($d_{\uparrow\downarrow}$) {EDM matrix elements of each of the individual molecules.}  
 {\it As a result, the flavor of the ED interaction  is completely determined by   single-molecule EDM matrix elements in the qubit basis, regardless of the precise nature of qubit states.}

  \begin{figure}[t]
    \includegraphics[width=0.7\columnwidth]{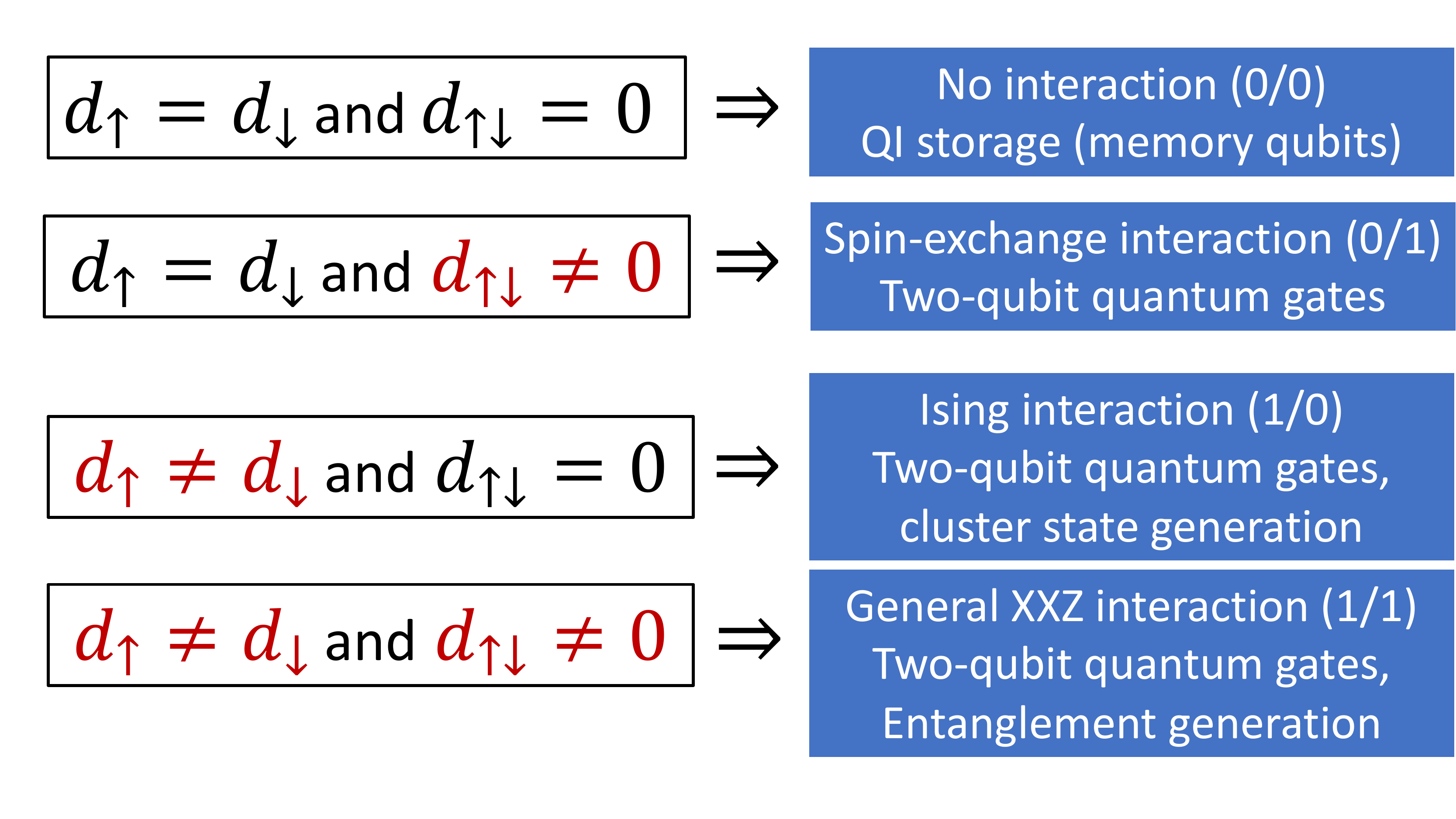}
    \caption{Molecular  qubit encoding classification diagram. To classify a qubit, first determine the values of $d_\uparrow$,  $d_\downarrow$, and $d_{\uparrow\downarrow}$, and  locate the relevant  box in the left column. The qubit encoding type is indicated to the right of each box. }
    \label{fig:diagram}
\end{figure}

It follows from Eq.~\eqref{eq:xxz_ham_main} that there can be 4 flavors of the ED interaction depending on whether or not the coupling constants $J_z$ and $J_\perp$ are equal to zero. If $J_z=J_\perp=0$, no ED interaction is present between the qubits, and we classify them as interactionless (0/0). In the later case, three types of the ED interaction can be distinguished based on whether the values of $|d_\uparrow-d_\downarrow|$ and $d_{\uparrow\downarrow}$  are zero (see Table 1). We classify these three types as Ising (1/0), spin-exchange (0/1), and  XXZ   (1/1).

 To identify the type of molecular qubit encoding, the reader can use the diagram shown in Fig.~\ref{fig:diagram}. One begins by calculating the matrix elements of the EDM in the qubit basis  $d_\uparrow, d_\downarrow$, and $d_{\uparrow\downarrow}$ and locating the corresponding box on the left-hand side of the diagram. The right-hand side of the diagram identifies the type of qubit encoding and outlines its possible applications, with a more detailed discussion provided in Sec. 2.

 To illustrate our classification scheme, we have considered a new type or molecular qubit encoding into non-adjacent rotational states of a polar diatomic molecule (such as $N=0$ and $N=2$). At zero external electric field, all EDM  matrix elements vanish due to the parity selection rules and this  encoding can be classified as interactionless (or 0/0 type), which can be  useful  as a storage (memory) qubit.
  In the presence of a dc electric field, the qubit is converted into type 1/1 due to both diagonal and off-diagonal EDM matrix elements being nonzero. However, when the projection $M_N$ of the rotational angular momentum of the qubit states differs by two or more, the off-diagonal  matrix elements of the EDM vanish identically, and the encoding type changes to 1/0.
  
  In future work, it would be interesting to apply our classification to identify novel encodings of molecular qubits, which could prove  useful for QIS applications. There remain pairs of molecular states, which have not been explored for qubit encoding, such as the  hyperfine components of different electronic and rovibrational states.
  % which could be potentially useful for various QIS applications. 
   It would also be interesting to extend our scheme to classify  qubit encodings in polyatomic molecules \cite{Yu:19, Albert:20}, which have recently been cooled and trapped in several laboratories \cite{Changala:19,Liu:22,Baum:20,Mitra:20}.

	\section*{Acknowledgements}
	We thank Ana Maria Rey for insightful discussions and for bringing to our attention the non-adjacent rotational state encoding. This work was supported by the NSF through the CAREER program (PHY-2045681). The work of KA and RVK is supported by NSERC of Canada.

	\section*{Appendix A: Effective electric dipole-dipole interaction}

	The ED interaction between molecules $i$ and $j$  takes the form \cite{Wall:15}
	\begin{align}
		H_{ij} =& R^{-3}  \left[ -\frac{3}{2} \sin^2\theta \left(\hat{d}_{1}\hat{d}_{1}e^{-2i\phi} + \rm{h.c.}\right)\right. \nonumber\\
		&\left. - \frac{3}{\sqrt{2}} \sin\theta \cos\theta \left[(\hat{d}_{1}\hat{d}_{0} + \hat{d}_{0}\hat{d}_{1})e^{-i\phi} + \rm{h.c.} \right] \right. \nonumber\\
		&\left. - \frac{1}{2}(3\cos^2\theta - 1) \left(2\hat{d}_{0}\hat{d}_{0} + \hat{d}_{1}\hat{d}_{-1} + \hat{d}_{-1}\hat{d}_{1}\right) \right]\label{eq:dipole_int_full}
	\end{align}
	where $\hat{d}_p$ are the spherical components of the dipole operator ($\hat{d}_0=\hat{d}_z$, $\hat{d}_{\pm1} = \mp(\hat{d}_x\pm i\hat{d}_y)/\sqrt{2}$), $R_{ij}$ is the distance between the molecules and $\theta_{ij}$ is the angle between the quantization vector and the vector connecting the two molecules. As the ED interaction is typically much weaker than the energy difference between the two qubit states $\ket{\uparrow}, \ket{\downarrow}$, processes that change the total magnetization are energetically off-resonant and can be neglected. With this assumption, the ED interaction can be described as the following spin Hamiltonian in the space spanned by $\{\ket{\uparrow\uparrow}, \ket{\uparrow\downarrow}, \ket{\downarrow\uparrow}, \ket{\downarrow\downarrow}\}$
	\begin{align}
		H_{ij} = \frac{1-3\cos^2\theta_{ij}}{R^3}&\left[\frac{J_\perp}{2}\left(S_i^+S_j^- + \rm{h.c.}\right) + J_zS_i^zS_j^z \right. \nonumber \\
		&\left.+ W\left(\mathbb{I}_iS_j^z + S_i^z\mathbb{I}_j\right) + V\mathbb{I}_i\mathbb{I}_j\vphantom{\frac{J_\perp}{2}}\right]. \label{eq:xxz_ham}
	\end{align}
	
	The dipole-dipole interaction of Eq. \eqref{eq:dipole_int_full} can be seen as a sum of terms that transfer $-2<q<2$ units of rotational angular momentum projection to the molecules' angular momentum projection. Assuming that states of different $M_N$ are not mixed together, terms with $q\neq 0$ are energetically off-resonant and Eq. \eqref{eq:dipole_int_full} can be simplified to only include terms with $q=0$:
	\begin{align}
		H_{ij} =& \frac{1 - 3\cos^2\theta_{ij}}{2R_{ij}^3} \left( 2\hat{d}_{0}\hat{d}_{0} + \hat{d}_{1}\hat{d}_{-1} + \hat{d}_{-1}\hat{d}_{1}\right). \label{eq:dipole_int}
	\end{align}
	
	Ignoring energetically off-resonant terms of $H_{ij}$, the interaction matrix can be written as
	\begin{align}
		\matr{H}_{ij} = \frac{1 - 3\cos^2 \theta}{R^3} \begin{pmatrix}
			H_{11} & 0 & 0 & 0 \\
			0 & H_{22} & H_{23} & 0 \\
			0 & H_{23}^* & H_{33} & 0 \\
			0 & 0 & 0 & H_{44}
		\end{pmatrix}.
	\end{align}
	To evaluate the couplings in Eq. \eqref{eq:xxz_ham}, we need to first the matrix elements of the interaction Hamiltonian. We define $d_\uparrow$, $d_\downarrow$, $d_{\uparrow\downarrow}$, $d^\pm_{\uparrow\downarrow}$ as the matrix elements of the dipole operators
	\begin{align}
		\braket{\uparrow|\hat{d}_0|\uparrow} &\equiv d_\uparrow, & \braket{\downarrow|\hat{d}_0|\downarrow}&\equiv d_\downarrow, \label{eq:dipoles1}\\ \braket{\uparrow|\hat{d}_0|\downarrow}&\equiv d_{\uparrow\downarrow}, & \braket{\uparrow|\hat{d}_{\pm 1}|\downarrow}&\equiv d^\pm_{\uparrow\downarrow}.\label{eq:dipoles2}
	\end{align}
	The matrix elements of $H_{ij}$, under the assumptions of Eq. \eqref{eq:dipole_int}, become
	\begin{align}
		H_{11} =&  \braket{\uparrow\uparrow|d_0d_0|\uparrow\uparrow} = d_\uparrow^2 \\
		H_{44} =&  \braket{\downarrow\downarrow|d_0d_0|\downarrow\downarrow} = d_\downarrow^2 \\
		H_{22} = H_{33} =&  \braket{\uparrow\downarrow|d_0d_0|\uparrow\downarrow} = d_\uparrow d_\downarrow \\
		H_{23} =& \braket{\uparrow\downarrow|d_0d_0|\downarrow\uparrow} \nonumber \\
		&+ \frac{1}{2} \left[ \braket{\uparrow\downarrow|d_1d_{-1}|\downarrow\uparrow} + \braket{\uparrow\downarrow|d_{-1}d_1|\downarrow\uparrow}\right] \nonumber \\
		=& d_{\uparrow\downarrow}^2 + \frac{1}{2} \left[d_{\uparrow\downarrow}^+d_{\downarrow\uparrow}^- + d_{\uparrow\downarrow}^-d_{\downarrow\uparrow}^+\right] \nonumber \\
		=& d_{\uparrow\downarrow}^2 - \frac{1}{2} \left[|d_{\uparrow\downarrow}^+|^2 + |d_{\uparrow\downarrow}^-|^2\right]
	\end{align}
	Since $\hat{d}_{\pm 1} = -\hat{d}_{\mp 1}^\dagger$, we get $d_{\uparrow\downarrow}^+ = -(d_{\downarrow\uparrow}^-)^*$ and $d_{\downarrow\uparrow}^+ = -(d_{\uparrow\downarrow}^-)^*$.
	
	The couplings of Eq. \eqref{eq:xxz_ham} are then derived as 
	\begin{align}
		J_z &= H_{11} - H_{22} - H_{33} + H_{44} \nonumber \\
		&= (d_\uparrow - d_\downarrow)^2, \label{eq:jz}\\
		J_\perp &= 2H_{23} \nonumber \\
		&= 2d_{\uparrow\downarrow}^2 - |d^+_{\uparrow\downarrow}|^2 - |d^-_{\uparrow\downarrow}|^2, \label{eq:jperp}\\
		W &= \frac{1}{2} (H_{11} - H_{44}) \nonumber \\
		&= (d_\uparrow^2 - d_\downarrow^2)/2, \label{eq:w}\\
		V &= \frac{1}{4}(H_{11} + H_{22} + H_{33} + H_{44}) \nonumber \\
		&= (d_\uparrow + d_\downarrow)^2/4, \label{eq:v}
	\end{align}
	in terms of the dipole matrix elements of Eqs. \eqref{eq:dipoles1} and \eqref{eq:dipoles2}.  Therefore, we observe that while all four terms depend on the matrix elements of the $\hat{d}_0$ operator, $J_\perp$ is also affected by the transition dipole matrix elements of $\hat{d}_{\pm 1}$. 
	
	\section*{Appendix B: Effective electric quadrupole-quadrupole interaction}
	
	The quadrupole-quadrupole interaction between molecules $i$ and $j$  takes the form 
	\begin{align}
		H_{ij} =& R^{-5} \left[\frac{35}{8}\sin^4\theta \left(q_2q_2e^{4i\phi} + \rm{h.c.}\right) + \frac{15}{4}\sqrt{7}\sin^3\theta\cos\theta\left[\left(q_2q_1 + q_1q_2\right)e^{-3i\phi} + \rm{h.c.}\right] \right. \nonumber\\
		& \left. + \frac{5}{4}\sin^2\theta(7\cos^2\theta - 1)\left[\left(\sqrt{\frac{3}{2}} q_2q_0 + \sqrt{\frac{3}{2}} q_0q_2 + 2q_1q_1\right) e^{2i\phi} + \rm{h.c.} \right] \right. \nonumber \\
		& \left. + \frac{15}{4}\sin\theta\cos\theta(\frac{7}{3}\cos^2\theta - 1)\left[\left(q_2q_{-1} + q_{-1}q_2 + \sqrt{6}q_1q_0 + q_0q_1\right)e^{i\phi} + \rm{h.c.}\right] \right. \nonumber \\
		& \left. + \frac{3}{8}(\frac{35}{3}\cos^4\theta - 10\cos^2\theta + 1)\left[\left(q_2q_{-2} + q_{-2}q_2 + 4(q_1q_{-1} + q_{-1}q_1) + 6q_0q_0\right)\right]\right] \label{eq:quadrupole_int_full}
	\end{align}
	where $\hat{q}_p$ are the spherical components of the quadrupole operator, $R_{ij}$ is the distance between the molecules and $\theta_{ij}$ is the angle between the quantization vector and the vector connecting the two molecules. 
	
	Using similar assumptions to \eqref{eq:dipole_int}, we can simplify \eqref{eq:quadrupole_int_full} to only include terms with $q=0$:
	\begin{align}
		H_{ij} =& R^{-5} \left[\frac{3}{8}(\frac{35}{3}\cos^4\theta - 10\cos^2\theta + 1)\left[\left(q_2q_{-2} + q_{-2}q_2 + 4(q_1q_{-1} + q_{-1}q_1) + 6q_0q_0\right)\right]\right]. \label{eq:quadrupole_int}
	\end{align}
	
	Ignoring energetically off-resonant terms of $H_{ij}$, the interaction matrix can be written as
	\begin{align}
		\matr{H}_{ij} = \frac{3}{8r^5}(\frac{35}{3}\cos^4\theta - 10\cos^2\theta + 1) \begin{pmatrix}
			H_{11} & 0 & 0 & 0 \\
			0 & H_{22} & H_{23} & 0 \\
			0 & H_{23}^* & H_{33} & 0 \\
			0 & 0 & 0 & H_{44}
		\end{pmatrix}.
	\end{align}
	We define $q_\uparrow, q_\downarrow, q_{\uparrow\downarrow}, q_{\uparrow\downarrow}^{\pm1}, q_{\uparrow\downarrow}^{\pm2}$ as the matrix elements of the quadrupole operators
	\begin{gather}
		\braket{\uparrow|\hat{q}_0|\uparrow} \equiv q_\uparrow, \quad \qquad \braket{\downarrow|\hat{q}_0|\downarrow}\equiv q_\downarrow \label{eq:quadrupoles1}\\ \braket{\uparrow|\hat{q}_0|\downarrow}\equiv q_{\uparrow\downarrow}, \qquad \braket{\uparrow|\hat{q}_{\pm 1}|\downarrow}\equiv q^{\pm1}_{\uparrow\downarrow} \label{eq:quadrupoles2} \\
		\braket{\uparrow|\hat{q}_{\pm 2}|\downarrow}\equiv q^{\pm2}_{\uparrow\downarrow}. \label{eq:quadrupoles3}
	\end{gather}
	
	The matrix elements of $H_{ij}$, under the assumptions of Eq. \eqref{eq:quadrupole_int}, become
	\begin{align}
		H_{11} =&  6\braket{\uparrow\uparrow|q_0q_0|\uparrow\uparrow} = 6q_\uparrow^2 \\
		H_{44} =&  6\braket{\downarrow\downarrow|q_0q_0|\downarrow\downarrow} = 6q_\downarrow^2 \\
		H_{22} = H_{33} =&  6\braket{\uparrow\downarrow|q_0q_0|\uparrow\downarrow} = 6q_\uparrow q_\downarrow \\
		H_{23} =& 6\braket{\uparrow\downarrow|q_0q_0|\downarrow\uparrow} \nonumber \\
		&+ 4\left[ \braket{\uparrow\downarrow|q_1q_{-1}|\downarrow\uparrow} + \braket{\uparrow\downarrow|q_{-1}q_1|\downarrow\uparrow}\right] \nonumber \\
		&+ \left[ \braket{\uparrow\downarrow|q_2q_{-2}|\downarrow\uparrow} + \braket{\uparrow\downarrow|q_{-2}q_2|\downarrow\uparrow}\right] \nonumber \\
		=& 6q_{\uparrow\downarrow}^2 + 4 \left[q_{\uparrow\downarrow}^{+1}q_{\downarrow\uparrow}^{-1} + q_{\uparrow\downarrow}^{-1}d_{\downarrow\uparrow}^{+1}\right] + \left[q_{\uparrow\downarrow}^{+2}q_{\downarrow\uparrow}^{-2} + q_{\uparrow\downarrow}^{-2}d_{\downarrow\uparrow}^{+2}\right] \nonumber \\
		=& 6q_{\uparrow\downarrow}^2 - 4 \left[|q_{\uparrow\downarrow}^{+1}|^2 + |q_{\uparrow\downarrow}^{-1}|^2\right] + \left[|q_{\uparrow\downarrow}^{+2}|^2 + |q_{\uparrow\downarrow}^{-2}|^2\right]
	\end{align}
	Since $\hat{q}_{\pm 1} = -\hat{q}_{\mp 1}^\dagger$ and $\hat{q}_{\pm 2} = \hat{q}_{\mp 2}^\dagger$, we get $q_{\uparrow\downarrow}^{+1} = -(q_{\downarrow\uparrow}^{-1})^*$,  $q_{\downarrow\uparrow}^{+1} = -(q_{\uparrow\downarrow}^{-1})^*$, $q_{\uparrow\downarrow}^{+2} = (q_{\downarrow\uparrow}^{-2})^*$ and  $q_{\downarrow\uparrow}^{+2} = (q_{\uparrow\downarrow}^{-2})^*$.
	
	The couplings of Eq. \eqref{eq:xxz_ham} are then derived as 
	\begin{align}
		J_z &= (q_\uparrow - q_\downarrow)^2, \label{eq:jz_q_app}\\
		J_\perp &= 2\left[6q_{\uparrow\downarrow}^2 - 4 \left[|q_{\uparrow\downarrow}^{+1}|^2 + |q_{\uparrow\downarrow}^{-1}|^2\right] + \left[|q_{\uparrow\downarrow}^{+2}|^2 + |q_{\uparrow\downarrow}^{-2}|^2\right]\right], \label{eq:jperp_q_app}\\
		W &= (q_\uparrow^2 - q_\downarrow^2)/2, \label{eq:w_q_app}\\
		V &= (q_\uparrow + q_\downarrow)^2/4, \label{eq:v_q_app}
	\end{align}
	in terms of the dipole matrix elements of Eqs. \eqref{eq:quadrupoles1}, \eqref{eq:quadrupoles2} and \eqref{eq:quadrupoles3}.  Therefore, we observe that while all four terms depend on the matrix elements of the $\hat{q}_0$ operator, $J_\perp$ is also affected by the transition dipole matrix elements of $\hat{q}_{\pm 1}$ and $\hat{q}_{\pm 2}$. 
	
	\bibliographystyle{unsrt}
	
	\bibliography{cold_mol.bib}

\providecommand{\latin}[1]{#1}
\makeatletter
\providecommand{\doi}
  {\begingroup\let\do\@makeother\dospecials
  \catcode`\{=1 \catcode`\}=2 \doi@aux}
\providecommand{\doi@aux}[1]{\endgroup\texttt{#1}}
\makeatother
\providecommand*\mcitethebibliography{\thebibliography}
\csname @ifundefined\endcsname{endmcitethebibliography}
  {\let\endmcitethebibliography\endthebibliography}{}
\begin{mcitethebibliography}{41}
\providecommand*\natexlab[1]{#1}
\providecommand*\mciteSetBstSublistMode[1]{}
\providecommand*\mciteSetBstMaxWidthForm[2]{}
\providecommand*\mciteBstWouldAddEndPuncttrue
  {\def\EndOfBibitem{\unskip.}}
\providecommand*\mciteBstWouldAddEndPunctfalse
  {\let\EndOfBibitem\relax}
\providecommand*\mciteSetBstMidEndSepPunct[3]{}
\providecommand*\mciteSetBstSublistLabelBeginEnd[3]{}
\providecommand*\EndOfBibitem{}
\mciteSetBstSublistMode{f}
\mciteSetBstMaxWidthForm{subitem}{(\alph{mcitesubitemcount})}
\mciteSetBstSublistLabelBeginEnd
  {\mcitemaxwidthsubitemform\space}
  {\relax}
  {\relax}

\bibitem[Bohn \latin{et~al.}(2017)Bohn, Rey, and Ye]{Bohn:17}
Bohn,~J.~L.; Rey,~A.~M.; Ye,~J. Cold molecules: Progress in quantum engineering
  of chemistry and quantum matter. \emph{Science} \textbf{2017}, \emph{357},
  1002--1010\relax
\mciteBstWouldAddEndPuncttrue
\mciteSetBstMidEndSepPunct{\mcitedefaultmidpunct}
{\mcitedefaultendpunct}{\mcitedefaultseppunct}\relax
\EndOfBibitem
\bibitem[Kaufman and Ni(2021)Kaufman, and Ni]{Kaufman:21}
Kaufman,~A.~M.; Ni,~K.-K. Quantum science with optical tweezer arrays of
  ultracold atoms and molecules. \emph{Nat. Phys.} \textbf{2021}, \emph{17},
  1324--1333\relax
\mciteBstWouldAddEndPuncttrue
\mciteSetBstMidEndSepPunct{\mcitedefaultmidpunct}
{\mcitedefaultendpunct}{\mcitedefaultseppunct}\relax
\EndOfBibitem
\bibitem[Sawant \latin{et~al.}(2020)Sawant, Blackmore, Gregory, Mur-Petit,
  Jaksch, Aldegunde, Hutson, Tarbutt, and Cornish]{Sawant:20}
Sawant,~R.; Blackmore,~J.~A.; Gregory,~P.~D.; Mur-Petit,~J.; Jaksch,~D.;
  Aldegunde,~J.; Hutson,~J.~M.; Tarbutt,~M.~R.; Cornish,~S.~L. Ultracold polar
  molecules as qudits. \emph{New J. Phys.} \textbf{2020}, \emph{22},
  013027\relax
\mciteBstWouldAddEndPuncttrue
\mciteSetBstMidEndSepPunct{\mcitedefaultmidpunct}
{\mcitedefaultendpunct}{\mcitedefaultseppunct}\relax
\EndOfBibitem
\bibitem[Albert \latin{et~al.}(2020)Albert, Covey, and Preskill]{Albert:20}
Albert,~V.~V.; Covey,~J.~P.; Preskill,~J. Robust Encoding of a Qubit in a
  Molecule. \emph{Phys. Rev. X} \textbf{2020}, \emph{10}, 031050\relax
\mciteBstWouldAddEndPuncttrue
\mciteSetBstMidEndSepPunct{\mcitedefaultmidpunct}
{\mcitedefaultendpunct}{\mcitedefaultseppunct}\relax
\EndOfBibitem
\bibitem[DeMille(2002)]{DeMille:02}
DeMille,~D. Quantum Computation with Trapped Polar Molecules. \emph{Phys. Rev.
  Lett.} \textbf{2002}, \emph{88}, 067901\relax
\mciteBstWouldAddEndPuncttrue
\mciteSetBstMidEndSepPunct{\mcitedefaultmidpunct}
{\mcitedefaultendpunct}{\mcitedefaultseppunct}\relax
\EndOfBibitem
\bibitem[Yelin \latin{et~al.}(2006)Yelin, Kirby, and C\^ot\'e]{Yelin:06}
Yelin,~S.~F.; Kirby,~K.; C\^ot\'e,~R. Schemes for robust quantum computation
  with polar molecules. \emph{Phys. Rev. A} \textbf{2006}, \emph{74},
  050301\relax
\mciteBstWouldAddEndPuncttrue
\mciteSetBstMidEndSepPunct{\mcitedefaultmidpunct}
{\mcitedefaultendpunct}{\mcitedefaultseppunct}\relax
\EndOfBibitem
\bibitem[Ni \latin{et~al.}(2018)Ni, Rosenband, and Grimes]{Ni:18}
Ni,~K.-K.; Rosenband,~T.; Grimes,~D.~D. Dipolar exchange quantum logic gate
  with polar molecules. \emph{Chem. Sci.} \textbf{2018}, \emph{9},
  6830--6838\relax
\mciteBstWouldAddEndPuncttrue
\mciteSetBstMidEndSepPunct{\mcitedefaultmidpunct}
{\mcitedefaultendpunct}{\mcitedefaultseppunct}\relax
\EndOfBibitem
\bibitem[Micheli \latin{et~al.}(2006)Micheli, Brennen, and Zoller]{Micheli:06}
Micheli,~A.; Brennen,~G.~K.; Zoller,~P. A toolbox for lattice-spin models with
  polar molecules. \emph{Nat. Phys.} \textbf{2006}, \emph{2}, 341--347\relax
\mciteBstWouldAddEndPuncttrue
\mciteSetBstMidEndSepPunct{\mcitedefaultmidpunct}
{\mcitedefaultendpunct}{\mcitedefaultseppunct}\relax
\EndOfBibitem
\bibitem[Bilitewski \latin{et~al.}(2021)Bilitewski, De~Marco, Li, Matsuda,
  Tobias, Valtolina, Ye, and Rey]{Bilitewski:21}
Bilitewski,~T.; De~Marco,~L.; Li,~J.-R.; Matsuda,~K.; Tobias,~W.~G.;
  Valtolina,~G.; Ye,~J.; Rey,~A.~M. Dynamical Generation of Spin Squeezing in
  Ultracold Dipolar Molecules. \emph{Phys. Rev. Lett.} \textbf{2021},
  \emph{126}, 113401\relax
\mciteBstWouldAddEndPuncttrue
\mciteSetBstMidEndSepPunct{\mcitedefaultmidpunct}
{\mcitedefaultendpunct}{\mcitedefaultseppunct}\relax
\EndOfBibitem
\bibitem[Tscherbul \latin{et~al.}(2023)Tscherbul, Ye, and Rey]{Tscherbul:23}
Tscherbul,~T.~V.; Ye,~J.; Rey,~A.~M. Robust Nuclear Spin Entanglement via
  Dipolar Interactions in Polar Molecules. \emph{Phys. Rev. Lett.}
  \textbf{2023}, \emph{130}, 143002\relax
\mciteBstWouldAddEndPuncttrue
\mciteSetBstMidEndSepPunct{\mcitedefaultmidpunct}
{\mcitedefaultendpunct}{\mcitedefaultseppunct}\relax
\EndOfBibitem
\bibitem[Park \latin{et~al.}(2017)Park, Yan, Loh, Will, and Zwierlein]{Park:17}
Park,~J.~W.; Yan,~Z.~Z.; Loh,~H.; Will,~S.~A.; Zwierlein,~M.~W. Second-scale
  nuclear spin coherence time of ultracold {$^{23}$Na$^{40}$K} molecules.
  \emph{Science} \textbf{2017}, \emph{357}, 372--375\relax
\mciteBstWouldAddEndPuncttrue
\mciteSetBstMidEndSepPunct{\mcitedefaultmidpunct}
{\mcitedefaultendpunct}{\mcitedefaultseppunct}\relax
\EndOfBibitem
\bibitem[Gregory \latin{et~al.}(2021)Gregory, Blackmore, Bromley, Hutson, and
  Cornish]{Gregory:21}
Gregory,~P.~D.; Blackmore,~J.~A.; Bromley,~S.~L.; Hutson,~J.~M.; Cornish,~S.~L.
  Robust storage qubits in ultracold polar molecules. \emph{Nat. Phys.}
  \textbf{2021}, \emph{17}, 1149--1153\relax
\mciteBstWouldAddEndPuncttrue
\mciteSetBstMidEndSepPunct{\mcitedefaultmidpunct}
{\mcitedefaultendpunct}{\mcitedefaultseppunct}\relax
\EndOfBibitem
\bibitem[Herrera \latin{et~al.}(2014)Herrera, Cao, Kais, and
  Whaley]{Herrera:14}
Herrera,~F.; Cao,~Y.; Kais,~S.; Whaley,~K.~B. Infrared-dressed entanglement of
  cold open-shell polar molecules for universal matchgate quantum computing.
  \emph{New J. Phys.} \textbf{2014}, \emph{16}, 075001\relax
\mciteBstWouldAddEndPuncttrue
\mciteSetBstMidEndSepPunct{\mcitedefaultmidpunct}
{\mcitedefaultendpunct}{\mcitedefaultseppunct}\relax
\EndOfBibitem
\bibitem[Karra \latin{et~al.}(2016)Karra, Sharma, Friedrich, Kais, and
  Herschbach]{Karra:16}
Karra,~M.; Sharma,~K.; Friedrich,~B.; Kais,~S.; Herschbach,~D. Prospects for
  quantum computing with an array of ultracold polar paramagnetic molecules.
  \emph{J. Chem. Phys.} \textbf{2016}, \emph{144}, 094301\relax
\mciteBstWouldAddEndPuncttrue
\mciteSetBstMidEndSepPunct{\mcitedefaultmidpunct}
{\mcitedefaultendpunct}{\mcitedefaultseppunct}\relax
\EndOfBibitem
\bibitem[Tesch and de~Vivie-Riedle(2002)Tesch, and de~Vivie-Riedle]{Tesch:02}
Tesch,~C.~M.; de~Vivie-Riedle,~R. Quantum Computation with Vibrationally
  Excited Molecules. \emph{Phys. Rev. Lett.} \textbf{2002}, \emph{89},
  157901\relax
\mciteBstWouldAddEndPuncttrue
\mciteSetBstMidEndSepPunct{\mcitedefaultmidpunct}
{\mcitedefaultendpunct}{\mcitedefaultseppunct}\relax
\EndOfBibitem
\bibitem[Zhao and Babikov(2006)Zhao, and Babikov]{Zhao:06}
Zhao,~M.; Babikov,~D. Phase control in the vibrational qubit. \emph{J. Chem.
  Phys.} \textbf{2006}, \emph{125}, 024105\relax
\mciteBstWouldAddEndPuncttrue
\mciteSetBstMidEndSepPunct{\mcitedefaultmidpunct}
{\mcitedefaultendpunct}{\mcitedefaultseppunct}\relax
\EndOfBibitem
\bibitem[Ospelkaus \latin{et~al.}(2010)Ospelkaus, Ni, Qu\'em\'ener, Neyenhuis,
  Wang, de~Miranda, Bohn, Ye, and Jin]{Ospelkaus:10a}
Ospelkaus,~S.; Ni,~K.-K.; Qu\'em\'ener,~G.; Neyenhuis,~B.; Wang,~D.;
  de~Miranda,~M. H.~G.; Bohn,~J.~L.; Ye,~J.; Jin,~D.~S. Controlling the
  Hyperfine State of Rovibronic Ground-State Polar Molecules. \emph{Phys. Rev.
  Lett.} \textbf{2010}, \emph{104}, 030402\relax
\mciteBstWouldAddEndPuncttrue
\mciteSetBstMidEndSepPunct{\mcitedefaultmidpunct}
{\mcitedefaultendpunct}{\mcitedefaultseppunct}\relax
\EndOfBibitem
\bibitem[Blackmore \latin{et~al.}(2020)Blackmore, Gregory, Bromley, and
  Cornish]{Blackmore:20b}
Blackmore,~J.~A.; Gregory,~P.~D.; Bromley,~S.~L.; Cornish,~S.~L. Coherent
  manipulation of the internal state of ultracold {$^{87}$Rb$^{133}$Cs}
  molecules with multiple microwave fields. \emph{Phys. Chem. Chem. Phys.}
  \textbf{2020}, \emph{22}, 27529--27538\relax
\mciteBstWouldAddEndPuncttrue
\mciteSetBstMidEndSepPunct{\mcitedefaultmidpunct}
{\mcitedefaultendpunct}{\mcitedefaultseppunct}\relax
\EndOfBibitem
\bibitem[Asnaashari and Krems(2022)Asnaashari, and Krems]{Asnaashari:22}
Asnaashari,~K.; Krems,~R.~V. Quantum annealing with pairs of
  ${}^{2}\mathrm{\ensuremath{\Sigma}}$ molecules as qubits. \emph{Phys. Rev. A}
  \textbf{2022}, \emph{106}, 022801\relax
\mciteBstWouldAddEndPuncttrue
\mciteSetBstMidEndSepPunct{\mcitedefaultmidpunct}
{\mcitedefaultendpunct}{\mcitedefaultseppunct}\relax
\EndOfBibitem
\bibitem[Nielsen and Chuang(2010)Nielsen, and Chuang]{Nielsen:10}
Nielsen,~M.~A.; Chuang,~I.~L. \emph{{\it Quantum Computation and Quantum
  Information: 10th Anniversary Edition}}; Cambridge University Press,
  2010\relax
\mciteBstWouldAddEndPuncttrue
\mciteSetBstMidEndSepPunct{\mcitedefaultmidpunct}
{\mcitedefaultendpunct}{\mcitedefaultseppunct}\relax
\EndOfBibitem
\bibitem[Lemeshko \latin{et~al.}(2013)Lemeshko, Krems, Doyle, and
  Kais]{Lemeshko:13}
Lemeshko,~M.; Krems,~R.~V.; Doyle,~J.~M.; Kais,~S. Manipulation of molecules
  with electromagnetic fields. \emph{Mol. Phys.} \textbf{2013}, \emph{111},
  1648--82\relax
\mciteBstWouldAddEndPuncttrue
\mciteSetBstMidEndSepPunct{\mcitedefaultmidpunct}
{\mcitedefaultendpunct}{\mcitedefaultseppunct}\relax
\EndOfBibitem
\bibitem[Wall \latin{et~al.}(2015)Wall, Hazzard, and Rey]{Wall:15c}
Wall,~M.~L.; Hazzard,~K. R.~A.; Rey,~A.~M. \emph{From Atomic to Mesoscale. The
  Role of Quantum Coherence in Systems of Various Complexities}; {World
  Scientific}, 2015; p~3\relax
\mciteBstWouldAddEndPuncttrue
\mciteSetBstMidEndSepPunct{\mcitedefaultmidpunct}
{\mcitedefaultendpunct}{\mcitedefaultseppunct}\relax
\EndOfBibitem
\bibitem[Gorshkov \latin{et~al.}(2011)Gorshkov, Manmana, Chen, Demler, Lukin,
  and Rey]{Gorshkov:11b}
Gorshkov,~A.~V.; Manmana,~S.~R.; Chen,~G.; Demler,~E.; Lukin,~M.~D.; Rey,~A.~M.
  Quantum magnetism with polar alkali-metal dimers. \emph{Phys. Rev. A}
  \textbf{2011}, \emph{84}, 033619\relax
\mciteBstWouldAddEndPuncttrue
\mciteSetBstMidEndSepPunct{\mcitedefaultmidpunct}
{\mcitedefaultendpunct}{\mcitedefaultseppunct}\relax
\EndOfBibitem
\bibitem[Aldegunde \latin{et~al.}(2008)Aldegunde, Rivington,
  \ifmmode~\dot{Z}\else \.{Z}\fi{}uchowski, and Hutson]{Aldegunde:08}
Aldegunde,~J.; Rivington,~B.~A.; \ifmmode~\dot{Z}\else
  \.{Z}\fi{}uchowski,~P.~S.; Hutson,~J.~M. Hyperfine energy levels of
  alkali-metal dimers: Ground-state polar molecules in electric and magnetic
  fields. \emph{Phys. Rev. A} \textbf{2008}, \emph{78}, 033434\relax
\mciteBstWouldAddEndPuncttrue
\mciteSetBstMidEndSepPunct{\mcitedefaultmidpunct}
{\mcitedefaultendpunct}{\mcitedefaultseppunct}\relax
\EndOfBibitem
\bibitem[Yan \latin{et~al.}(2013)Yan, Moses, Gadway, Covey, Hazzard, Rey, Jin,
  and Ye]{Yan:13}
Yan,~B.; Moses,~S.~A.; Gadway,~B.; Covey,~J.~P.; Hazzard,~K. R.~A.; Rey,~A.~M.;
  Jin,~D.~S.; Ye,~J. Observation of dipolar spin-exchange interactions with
  lattice-confined polar molecules. \emph{Nature} \textbf{2013}, \emph{501},
  521--525\relax
\mciteBstWouldAddEndPuncttrue
\mciteSetBstMidEndSepPunct{\mcitedefaultmidpunct}
{\mcitedefaultendpunct}{\mcitedefaultseppunct}\relax
\EndOfBibitem
\bibitem[Lin \latin{et~al.}(2022)Lin, He, Jin, Chen, and Wang]{Lin:22}
Lin,~J.; He,~J.; Jin,~M.; Chen,~G.; Wang,~D. Seconds-Scale Coherence on Nuclear
  Spin Transitions of Ultracold Polar Molecules in 3D Optical Lattices.
  \emph{Phys. Rev. Lett.} \textbf{2022}, \emph{128}, 223201\relax
\mciteBstWouldAddEndPuncttrue
\mciteSetBstMidEndSepPunct{\mcitedefaultmidpunct}
{\mcitedefaultendpunct}{\mcitedefaultseppunct}\relax
\EndOfBibitem
\bibitem[Gorshkov \latin{et~al.}(2011)Gorshkov, Manmana, Chen, Ye, Demler,
  Lukin, and Rey]{Gorshkov:11a}
Gorshkov,~A.~V.; Manmana,~S.~R.; Chen,~G.; Ye,~J.; Demler,~E.; Lukin,~M.~D.;
  Rey,~A.~M. Tunable Superfluidity and Quantum Magnetism with Ultracold Polar
  Molecules. \emph{Phys. Rev. Lett.} \textbf{2011}, \emph{107}, 115301\relax
\mciteBstWouldAddEndPuncttrue
\mciteSetBstMidEndSepPunct{\mcitedefaultmidpunct}
{\mcitedefaultendpunct}{\mcitedefaultseppunct}\relax
\EndOfBibitem
\bibitem[Briegel and Raussendorf(2001)Briegel, and Raussendorf]{Briegel:01}
Briegel,~H.~J.; Raussendorf,~R. Persistent Entanglement in Arrays of
  Interacting Particles. \emph{Phys. Rev. Lett.} \textbf{2001}, \emph{86},
  910--913\relax
\mciteBstWouldAddEndPuncttrue
\mciteSetBstMidEndSepPunct{\mcitedefaultmidpunct}
{\mcitedefaultendpunct}{\mcitedefaultseppunct}\relax
\EndOfBibitem
\bibitem[Briegel \latin{et~al.}(2009)Briegel, Browne, D{\"u}r, Raussendorf, and
  Van~den Nest]{Briegel:09}
Briegel,~H.~J.; Browne,~D.~E.; D{\"u}r,~W.; Raussendorf,~R.; Van~den Nest,~M.
  Measurement-based quantum computation. \emph{Nat. Phys.} \textbf{2009},
  \emph{5}, 19--26\relax
\mciteBstWouldAddEndPuncttrue
\mciteSetBstMidEndSepPunct{\mcitedefaultmidpunct}
{\mcitedefaultendpunct}{\mcitedefaultseppunct}\relax
\EndOfBibitem
\bibitem[Jones(2003)]{Jones:02}
Jones,~J.~A. Robust Ising gates for practical quantum computation. \emph{Phys.
  Rev. A} \textbf{2003}, \emph{67}, 012317\relax
\mciteBstWouldAddEndPuncttrue
\mciteSetBstMidEndSepPunct{\mcitedefaultmidpunct}
{\mcitedefaultendpunct}{\mcitedefaultseppunct}\relax
\EndOfBibitem
\bibitem[Ma \latin{et~al.}(2011)Ma, Wang, Sun, and Nori]{Ma:11}
Ma,~J.; Wang,~X.; Sun,~C.~P.; Nori,~F. Quantum spin squeezing. \emph{Phys.
  Rep.} \textbf{2011}, \emph{509}, 89--165\relax
\mciteBstWouldAddEndPuncttrue
\mciteSetBstMidEndSepPunct{\mcitedefaultmidpunct}
{\mcitedefaultendpunct}{\mcitedefaultseppunct}\relax
\EndOfBibitem
\bibitem[Pezz\`e \latin{et~al.}(2018)Pezz\`e, Smerzi, Oberthaler, Schmied, and
  Treutlein]{Pezze:18}
Pezz\`e,~L.; Smerzi,~A.; Oberthaler,~M.~K.; Schmied,~R.; Treutlein,~P. Quantum
  metrology with nonclassical states of atomic ensembles. \emph{Rev. Mod.
  Phys.} \textbf{2018}, \emph{90}, 035005\relax
\mciteBstWouldAddEndPuncttrue
\mciteSetBstMidEndSepPunct{\mcitedefaultmidpunct}
{\mcitedefaultendpunct}{\mcitedefaultseppunct}\relax
\EndOfBibitem
\bibitem[Rost \latin{et~al.}(1992)Rost, Griffin, Friedrich, and
  Herschbach]{Rost:92}
Rost,~J.~M.; Griffin,~J.~C.; Friedrich,~B.; Herschbach,~D.~R. Pendular states
  and spectra of oriented linear molecules. \emph{Phys. Rev. Lett.}
  \textbf{1992}, \emph{68}, 1299--1302\relax
\mciteBstWouldAddEndPuncttrue
\mciteSetBstMidEndSepPunct{\mcitedefaultmidpunct}
{\mcitedefaultendpunct}{\mcitedefaultseppunct}\relax
\EndOfBibitem
\bibitem[Bhongale \latin{et~al.}(2013)Bhongale, Mathey, Zhao, Yelin, and
  Lemeshko]{Bhongale:13}
Bhongale,~S.~G.; Mathey,~L.; Zhao,~E.; Yelin,~S.~F.; Lemeshko,~M. Quantum
  Phases of Quadrupolar Fermi Gases in Optical Lattices. \emph{Phys. Rev.
  Lett.} \textbf{2013}, \emph{110}, 155301\relax
\mciteBstWouldAddEndPuncttrue
\mciteSetBstMidEndSepPunct{\mcitedefaultmidpunct}
{\mcitedefaultendpunct}{\mcitedefaultseppunct}\relax
\EndOfBibitem
\bibitem[Yu \latin{et~al.}(2019)Yu, Cheuk, Kozyryev, and Doyle]{Yu:19}
Yu,~P.; Cheuk,~L.~W.; Kozyryev,~I.; Doyle,~J.~M. A scalable quantum computing
  platform using symmetric-top molecules. \emph{New J. Phys.} \textbf{2019},
  \emph{21}, 093049\relax
\mciteBstWouldAddEndPuncttrue
\mciteSetBstMidEndSepPunct{\mcitedefaultmidpunct}
{\mcitedefaultendpunct}{\mcitedefaultseppunct}\relax
\EndOfBibitem
\bibitem[Changala \latin{et~al.}(2019)Changala, Weichman, Lee, Fermann, and
  Ye]{Changala:19}
Changala,~P.~B.; Weichman,~M.~L.; Lee,~K.~F.; Fermann,~M.~E.; Ye,~J.
  Rovibrational quantum state resolution of the {C$_{60}$} fullerene.
  \emph{Science} \textbf{2019}, \emph{363}, 49--54\relax
\mciteBstWouldAddEndPuncttrue
\mciteSetBstMidEndSepPunct{\mcitedefaultmidpunct}
{\mcitedefaultendpunct}{\mcitedefaultseppunct}\relax
\EndOfBibitem
\bibitem[Liu \latin{et~al.}(2022)Liu, Changala, Weichman, Liang, Toscano,
  K\l{}os, Kotochigova, Nesbitt, and Ye]{Liu:22}
Liu,~L.~R.; Changala,~P.~B.; Weichman,~M.~L.; Liang,~Q.; Toscano,~J.;
  K\l{}os,~J.; Kotochigova,~S.; Nesbitt,~D.~J.; Ye,~J. Collision-Induced
  ${\mathrm{C}}_{60}$ Rovibrational Relaxation Probed by State-Resolved
  Nonlinear Spectroscopy. \emph{PRX Quantum} \textbf{2022}, \emph{3},
  030332\relax
\mciteBstWouldAddEndPuncttrue
\mciteSetBstMidEndSepPunct{\mcitedefaultmidpunct}
{\mcitedefaultendpunct}{\mcitedefaultseppunct}\relax
\EndOfBibitem
\bibitem[Baum \latin{et~al.}(2020)Baum, Vilas, Hallas, Augenbraun, Raval,
  Mitra, and Doyle]{Baum:20}
Baum,~L.; Vilas,~N.~B.; Hallas,~C.; Augenbraun,~B.~L.; Raval,~S.; Mitra,~D.;
  Doyle,~J.~M. 1D Magneto-Optical Trap of Polyatomic Molecules. \emph{Phys.
  Rev. Lett.} \textbf{2020}, \emph{124}, 133201\relax
\mciteBstWouldAddEndPuncttrue
\mciteSetBstMidEndSepPunct{\mcitedefaultmidpunct}
{\mcitedefaultendpunct}{\mcitedefaultseppunct}\relax
\EndOfBibitem
\bibitem[Mitra \latin{et~al.}(2020)Mitra, Vilas, Hallas, Anderegg, Augenbraun,
  Baum, Miller, Raval, and Doyle]{Mitra:20}
Mitra,~D.; Vilas,~N.~B.; Hallas,~C.; Anderegg,~L.; Augenbraun,~B.~L.; Baum,~L.;
  Miller,~C.; Raval,~S.; Doyle,~J.~M. Direct laser cooling of a symmetric top
  molecule. \emph{Science} \textbf{2020}, \emph{369}, 1366--1369\relax
\mciteBstWouldAddEndPuncttrue
\mciteSetBstMidEndSepPunct{\mcitedefaultmidpunct}
{\mcitedefaultendpunct}{\mcitedefaultseppunct}\relax
\EndOfBibitem
\bibitem[Wall \latin{et~al.}(2015)Wall, Maeda, and Carr]{Wall:15}
Wall,~M.~L.; Maeda,~K.; Carr,~L.~D. Realizing unconventional quantum magnetism
  with symmetric top molecules. \emph{New J. Phys.} \textbf{2015}, \emph{17},
  025001\relax
\mciteBstWouldAddEndPuncttrue
\mciteSetBstMidEndSepPunct{\mcitedefaultmidpunct}
{\mcitedefaultendpunct}{\mcitedefaultseppunct}\relax
\EndOfBibitem
\end{mcitethebibliography}

\end{document}